\documentclass[aps,prb,twocolumn,groupedaddress,showpacs,superscriptaddress,amssymb,amsmath,noeprint]{revtex4-2}
\usepackage{graphicx}
\usepackage[english]{babel}
\usepackage{dcolumn}
\usepackage{bm}
\usepackage{hyperref}
\usepackage{cleveref}
\usepackage{multirow}
\hypersetup{
    colorlinks=true,
    linkcolor=blue,
    urlcolor=blue,
    citecolor=blue
}
\usepackage{xcolor}
\usepackage{soul}
\usepackage{float}
\definecolor{sarkar}{rgb}{0.0588,0.3216,0.7294}
\definecolor{bijay}{rgb}{1,0.4549,0.0902}
\definecolor{devendra}{rgb}{1,0,0.5647}
\definecolor{lokesh}{rgb}{0.5059,0.8471,0.8157}

\newcommand{\tr}{\mathrm{Tr}}

\newcommand{\df}[1]{\textcolor{black}{#1}}
\begin{document}
\newcolumntype{M}[1]{>{\centering\arraybackslash}m{#1}}

\title{Bipartite particle number fluctuations in dephased long-range lattice systems}

\author{Lokesh Tater}
\email{lokesh.tater@students.iiserpune.ac.in} 
\affiliation{Department of Physics, Indian Institute of Science Education and Research, Pune 411008, India}

\author{Subhajit Sarkar}
\email{sbhjt72@gmail.com}
\affiliation{Department of Physics and Nanotechnology, SRM Institute of Science and Technology, Kattankulathur-603 203, India}

\author{Devendra Singh Bhakuni}
\email{dbhakuni@ictp.it}
\affiliation{The Abdus Salam International Centre for Theoretical Physics, Strada Costiera 11, 34151 Trieste, Italy}

\author{Bijay Kumar Agarwalla}
\email{bijay@iiserpune.ac.in}
\affiliation{Department of Physics, Indian Institute of Science Education and Research, Pune 411008, India}

\date{\today} 

\begin{abstract} 
\df{We investigate the dynamics of subsystem particle number fluctuations in a long-range system with power-law decaying hopping strength characterized by exponent $\mu$ and subjected to a local dephasing at every site. We introduce an efficient {\it bond length} representation for the four-point correlator, enabling the large-scale simulation of the dynamics of particle number fluctuations from translationally invariant initial states. Our results show that the particle number fluctuation dynamics exhibit one-parameter Family-Vicsek scaling, with superdiffusive scaling exponents for $\mu < 1.5$ and diffusive scaling exponents for $\mu \geq 1.5$. Finally, exploiting the bond-length representation, we provide an exact analytical expression for the particle number fluctuations and their scaling exponents in the short-range limit ($\mu \to \infty)$.}
\end{abstract}

\maketitle 

{\it Introduction.--} \df{The universal properties of nonequilibrium dynamics in many-body quantum systems, particularly the relaxation of both local and non-local observables, have attracted significant attention in recent years \cite{abanin2019many,deutsch1991quantum,Deutsch2018eth,polkovnikov2011noneq,srednicki1994chaos,KPZ_heisenberg_magnet,Ueda2020, Rigol2008,wienand2024emergence,hopjan2023scale,Fujimoto_exact,Torres2015relaxtaion,annurev_Huse,Eisert2015quantum,gogolin2016equilibration}. The eigenstate thermalization hypothesis (ETH), for example, successfully predicts the rapid relaxation of local observables \cite{abanin2019many, Deutsch2018eth, D'Alessio03052016, srednicki1994chaos}. In contrast, the relaxation dynamics of non-local observables extend beyond ETH and their equilibration is described by a few-parameter hydrodynamic model, reflecting the significantly slower relaxation processes involved \cite{Sommer2011universal,Erne2018universal,ruggiero2020quantum,doyon2017large,castro2016emergent,bulchandani2018bethe,lux2014hydrodynamic,capizzi2025hydrodynamics}.
Furthermore, advances in experimental techniques, particularly the ability of quantum simulators to achieve single-site resolution, have enabled access to different non-local observables \cite{Gross2021,Rispoli2019,sherson2010,joshi2022observing,impertro2023local,Bernien2017,Bakr2009,Christakis2023,wienand2024emergence,KPZ_heisenberg_magnet,rosenberg2024dynamics,browaeys2020many,joshi2025measuring,mark2024maximum}.}

\df{A key non-local observable in quantum systems is the subsystem particle number fluctuation, which quantifies the variance of particle densities within a subsystem~\cite{Klich2009quantum,gioev2006entanglement,song2010general,Rachel2012detecting,agrawal2022entanglment,oshima2023charge,lukin2019probing,lukin2019probing,kiefer2020bounds,kiefer2021unlimited,luitz2020absence,ghosh2022resonance,singh2016signatures,serbyn2015criterion,aceituno2024ultraslow,armijo2010probing,Collura_2017,levine2012full,Calabrese_2012,sarang2024distinct}. This measure has recently been used in experiments to understand the emergence of fluctuating hydrodynamics in chaotic quantum systems~\cite{wienand2024emergence} and has subsequently been understood theoretically~\cite{Fujimoto_exact,gamayun2024fullcounting}. Beyond this, bipartite particle fluctuation has also appeared in other significant contexts, including quantum surface roughening dynamics \cite{Fujimoto_dynamical,Fujimoto_FV_Bose_Gas,SR_Sreemayee,bhakuni2024dynamic}, its connection to entanglement entropy \cite{Klich2009quantum,gioev2006entanglement,song2010general,lukin2019probing}, many-body localization~\cite{lukin2019probing,kiefer2020bounds,kiefer2021unlimited,luitz2020absence,ghosh2022resonance,singh2016signatures,serbyn2015criterion,aceituno2024ultraslow}, and the characterization of phase transitions~\cite{Rachel2012detecting,agrawal2022entanglment,oshima2023charge}, among others. 
Recently, for isolated systems, the spatio-temporal growth of quantum surface roughness --representing particle number fluctuations over a given length scale-- and its subsequent saturation has been shown to follow the Family-Vicsek dynamical scaling \cite{family_scaling_1985,vicsek_dynamic_1984,vicsek_self-affine_1990}, with the corresponding growth exponents providing information about the underlying universality class \cite{Fujimoto_FV_Bose_Gas,Fujimoto_dynamical,Fujimoto_impact,bhakuni2024dynamic,SR_Sreemayee}.
}

As quantum systems inevitably interact with their environment, leading to dissipation and decoherence, the dynamics of local and non-local observables, along with their corresponding scaling behaviours, can be significantly affected \cite{Yin_scaling_critical_PhysRevB_2016, Rossini_PhysRevB_scaling_ness_2019, gleis2024dynamicalscalingplanckiandissipation, kou2025kibblezurekscalingimmuneantikibblezurek, Liu_PhysRevB_universal_KPZ_2023,bhakuni2024noise}. For instance, a recent study based on the dynamics of particle number fluctuation or surface roughness growth has shown that dissipation can alter the universality class from ballistic scaling to Edwards-Wilkinson scaling with diffusive transport \cite{Fujimoto_impact}. However, research in this
direction has so far focused primarily on lattice systems with short-range hopping and interactions \cite{Fujimoto_exact, Fujimoto_dynamical, Fujimoto_impact, Fujimoto_FV_Bose_Gas}, leaving open questions about the dynamics of bipartite particle fluctuations in long-range systems with power-law hopping and, moreover, with environmental interactions.  

Long-range systems are prevalent in nature, as well as in artificial light-harvesting systems \cite{long_range_natural1,long_range_natural2,long_range_artifical1,long_range_artifical2,long_range_artifical3}, and can be engineered in experimental setups~\cite{Blatt2012, Jurcevic2014,Zhang2017,monroe2021programmable,joshi2022observing,browaeys2020many,muller2012engineered,joshi2020quantum,Scholl2021,Britton2012,porras2004effective}. They exhibit significant qualitative changes in equilibrium phases, ground state properties, and dynamics \cite{defenu2023long,Kosterlitz_PhysRevLett, kuwahara2020area, koffel2012entanglement, vodola2015long, chen2019finite, tran2020hierarchy, Kuwahara_light_cones2020, Zhou_levy_flight2020, Kuwahara_noscrambling2020, Sondhi_MBL_long_PhysRevX,Prosen_long_range,bhakuni2021suppression} leading to rich and unconventional physics. This includes measurement-induced entanglement transitions \cite{block_measurement-induced_2022, muller_measurement-induced_2022, Minato_MIPT_long_range}, transitions from normal to anomalous transport in steady-state currents under bulk dephasing \cite{Subhajit_long_range, Abhinav_long_range}, among other phenomena.

In this Letter, we consider a long-range fermionic lattice system that is subjected to number-conserving dephasing probes at each of its sites and investigate the growth and subsequent thermalization of bipartite particle number fluctuation starting from alternating and domain-wall initial state. We focus on the strong dephasing limit that allows us to obtain effective dynamics for two and four-point correlators, which are required to compute the number fluctuation in a domain. For alternating initial state, we device a novel bond-length representation for the four-point correlators that enables us to simulate number fluctuation for significantly large system size ($\approx 10^4$ sites) which is otherwise severely limited. We further provide exact analytical results for the scaling exponents for the dephased lattice in the short-range limit. 

{\it Long-range lattice setup and bipartite number fluctuation.--} We consider a one-dimensional long-range free-fermionic lattice consisting of $L$ sites with hopping between sites decaying as a power-law with distance. The Hamiltonian of the lattice is written as \cite{Prosen_long_range, Subhajit_long_range, Abhinav_long_range, Knap_nonlocal}
\begin{equation}
\label{hamiltonian}
\hat{H} = J\sum_{i=0}^{L-1} \sum_{j>i}^{L-1} \frac{1}{d(i,j)^\mu}\Big[\hat{c}_i\hat{c}_{j}^{\dagger} + \hat{c}_i^{\dagger}\hat{c}_j\Big],
\end{equation}
where $J$ is the hopping strength and $\mu$ is the power-law hopping exponent, which is always chosen to be larger than 1 here. The function $d(i,j)$ denotes the distance between lattice sites $i$ and $j$, and for the periodic boundary condition, it is given as $d(i,j)={\rm min}(|i-j|, L-|i-j|)$. The operators $\hat{c}_i$ ($\hat{c}^{\dagger}_i$) are the fermionic annihilation (creation) operators for site $i$. Throughout this work, we set $\hbar=1$.

\df{We further consider a particle number-conserving dephasing probe at each site of the lattice \cite{Giamarchi_transport2024, Subhajit_long_range,Abhinav_long_range,Prosen_long_range,Demler, PhysRevLett.118.140403,Ghosh_2024,Goold2021,Landi_review,Longhi2024,liang2024dephasing,Znidaric_2010,Fujimoto_impact}. Such a local dephasing can arise from thermal noise or local coupling to the vibrational degrees of freedom~\cite{mukamel1978nature}. The dynamics of the setup can be modelled using the Lindblad quantum master equation (LQME) \cite{lindblad1976generators},}
\begin{equation}
\label{eq:master_eq}
    \frac{d}{dt} \hat{\rho} = -i\Big[\hat{H},{\hat{\rho}(t)}\Big] + \gamma\,\sum_{j=0}^{L-1} \Big(\hat{n}_j\,\hat{\rho}\,\hat{n}_j -\frac{1}{2}\{\hat{n}_j,\hat{\rho}\} \Big),
\end{equation}
where the first term represents that standard unitary contribution and the second term represents the Lindblad dissipator corresponding to on-site dephasing. Here $\hat{n}_j=\hat{c}_j^{\dagger} \hat{c}_j$ is the fermion number operator at site $j$ and $\gamma$ is the dephasing strength.

Given the setup, we study the dynamics of particle number fluctuation within a subsystem of size $L/2$ (considering $L$ to be even here). To start with, we focus on the alternating initial state where every even site of the lattice is initially occupied. The density matrix corresponding to this state is given by $\hat{\rho}_{\mathrm{alt}}=|\psi_{\mathrm{alt}}\rangle\langle \psi_{\mathrm{alt}}|$ with $|\psi_{\mathrm{alt}}\rangle=\displaystyle \prod_{m=0}^{L/2-1}\hat{c}^{\dagger}_{2m} |0 \rangle$.
For such an initial state, we further provide analytical results for particle number fluctuation in some specific limits.

In the supplemental material~\cite{supp}, we analyse the dynamics starting with the domain-wall initial state and compare it with that of the alternating state. We highlight both qualitative similarities and key differences in their temporal evolution.

The particle number fluctuation in a domain of size $L/2$ of a lattice is defined as $\sigma^2(L,t)=\langle{\hat{N}_{L/2}}^2\rangle-\langle{\hat{N}_{L/2}}\rangle^2$ where $\hat{N}_{L/2}= \displaystyle \sum_{i=0}^{L/2-1}\hat{n}_i$ gives the number of particles in the left half of the lattice. The expectation value $\langle * \rangle = {\rm Tr}\big[* \hat{\rho}(t)\big]$ is taken with respect to the time-evolved density matrix $\hat{\rho}(t)$ which evolves according to the LQME in Eq.~\eqref{eq:master_eq}. In particular, recent studies have discovered an interesting link between such particle number fluctuation and surface roughness in quantum systems \cite{Fujimoto_impact,Fujimoto_dynamical,Fujimoto_FV_Bose_Gas,Fujimoto_exact}. The number fluctuation $\sigma^2(L,t)$ can be expressed in terms of two-point $D_{mn}=\langle\hat{c}^{\dagger}_m \hat{c}_n\rangle$, and four-point correlators $F_{mnpq}=\langle\hat{c}^{\dagger}_m \hat{c}^{\dagger}_n \hat{c}_p\hat{c}_q\rangle$ as 
\begin{equation}
    \sigma^2(L,t) = -\sum_{m,n=0}^{L/2-1}F_{mnmn} + \sum_{m=0}^{L/2-1}D_{mm} \Big[1 -\sum_{n=0}^{L/2-1} D_{nn}\Big].
    \label{number-fluc}
\end{equation}
It is important to note that, due to the presence of the quartic term in the Lindblad dissipator in Eq.~\eqref{eq:master_eq}, Wick's theorem does not hold in this setup. As a result, four-point correlators cannot be decomposed into products of two-point correlators and must be evaluated independently. 

The LQME given in Eq.~\eqref{eq:master_eq} conserves the total particle number, thereby confining the dynamics to a fixed particle number sector. Additionally, the jump operator $\hat{n}_i$ being Hermitian, the equation of motion for $n-$ point correlator depends only on correlators of order $n$ or lower, without involving the higher order ones \cite{Demler}. Thus we focus on equations of motion of two-point and four-point correlators and restrict our analysis to the strong dephasing regime, $\gamma \gg J$. The strong dephasing limit is particularly interesting, as it leads to the loss of quantum coherence while still permitting the non-trivial dynamics of particle number fluctuations. This regime often serves as a bridge between quantum and classical behaviour and offers a tractable setting -- both numerically and analytically -- for exploring the intricate interplay between dissipation and long-range hopping. 

\begin{figure*}[t]
    \centering
    \includegraphics[width=\linewidth]{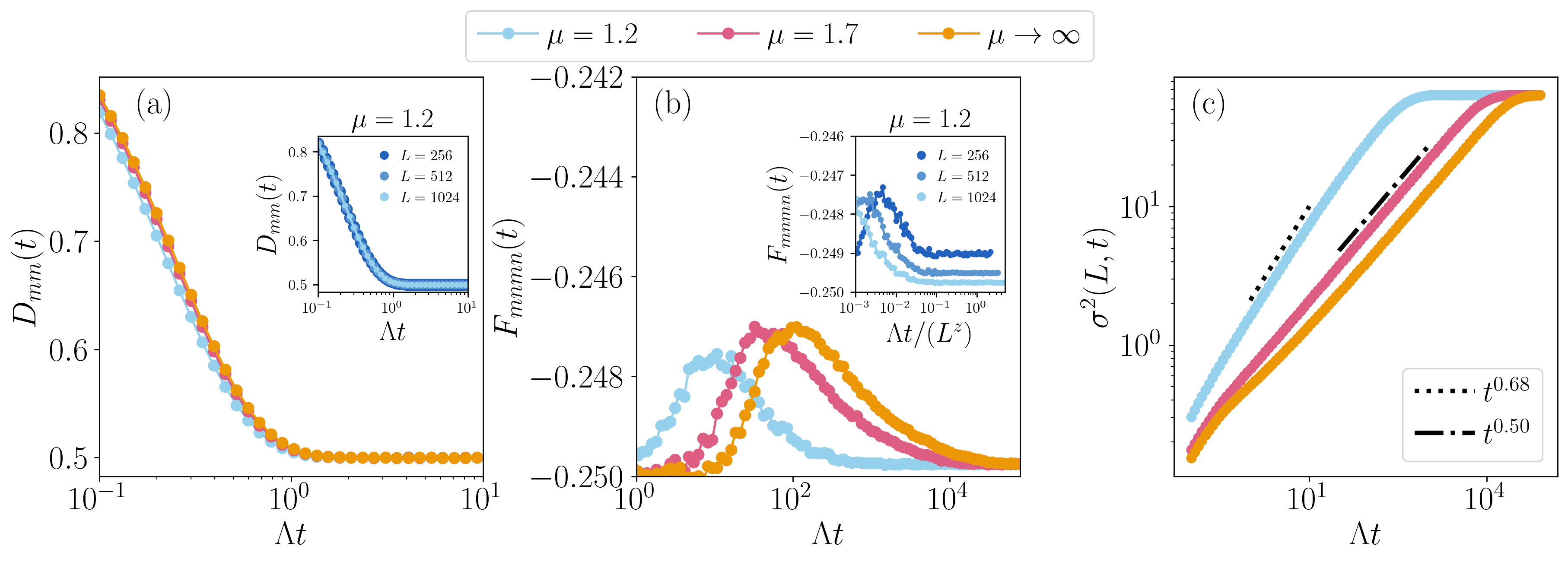}
    \caption{Plot for the dynamics of (a) diagonal of the one-point correlator $D_{mm}(t)$, i.e., the local density $n_m(t)$ for $m=0$, (b) the diagonal of four-point correlators $F_{mnmn}(t)$ for $(m,n)=(0,21)$, and (c) the particle number fluctuation $\sigma^2(L,t)$ in a subsystem of size $L/2$, for different values of long-range hopping $\mu$ starting with an alternating initial state. The system size is taken to be $L=1024$. The dotted and the dashed line indicates the exponent values $\beta$ for $\sigma^2(L,t) \sim t^{\beta}$ for different $\mu$ values.
    The inset in (a) shows system size-independent relaxation of local density. The inset of (b) shows the system size-dependent saturation time scale for the four-point correlator by re-scaling the $x$-axis after dividing by $L^z$ with $z=2\mu -1$ and $\mu=1.2$. To show this scaling collapse, we choose three different system size values with $L=256$, $L=512$, and $L=1024$.}
    \label{fig:dynamics_one_two}
\end{figure*}

{\it Effective dynamical equations for two and four-point correlators.--} In the strong dephasing limit, the system achieves decoherence much before equilibrating to a maximally mixed state. Hence, late-time dynamics ($t\gg1/\gamma$) can be mapped to a classical Markov process involving only the populations of the classical states. Following Ref.~\cite{Prosen_long_range}, we adiabatically eliminate the off-diagonal (coherence) elements of the one-particle density matrix, i.e., two-point correlation function $D_{mn}$ with $m\neq n$ to obtain the effective equation for the diagonals $D_{mm}$ (see Supplemental material \cite{supp} for the details):
\begin{equation}
\label{eq:2pt_effective}
    \partial_t D_{mm} = \Lambda \sum_{j = \pm 1}^{\pm (\frac{L}{2}-1), \frac{L}{2}} \frac{D_{(m+j)(m+j)} - D_{mm}}{|j|^{2\mu}},
\end{equation}
where $(m+j)\equiv (m+j)\mod L$, and $\Lambda=2J^2/\gamma$ can be interpreted as the diffusion constant. Following a similar approach, the adiabatic elimination of coherences 
for the two-particle density matrix i.e., four-point correlation function $F_{mnpq}$ yields the following effective equation for $F_{mnmn}$:
\begin{eqnarray}
\label{eq:4pt_effective}
\partial_t F_{mnmn}&=&(1-\delta^m_n)\times \nonumber \\
\Lambda &&\!\!\!\!\!\!\!\sum_{j = \pm 1}^{\pm (\frac{L}{2}-1), \frac{L}{2}}\bigg( (1-\delta^m_{n+j})\,\frac{F_{m(n+j)m(n+j)}-F_{mnmn}}{|j|^{2\mu}} \nonumber \\   
&+& (1-\delta^{m+j}_{n})\,\frac{ F_{(m+j)n(m+j)n}-F_{mnmn}}{|j|^{2\mu}} \bigg),\,\,
\end{eqnarray}
where $(m+j)\equiv (m+j)\mod L$, $(n+j)\equiv (n+j)\mod L$ and $\delta^i_j$ is the Kronecker delta. The presence of the terms $(1-\delta^i_j)$ in Eq.~(\ref{eq:4pt_effective}) is a consequence of fermionic statistics, which ensures that the diagonals of the form $F_{mmmm}$ vanish. Note that while deriving Eq.~\eqref{eq:2pt_effective} and Eq.~\eqref{eq:4pt_effective} we assume periodic boundary conditions, and hence the summation $j$ in is over the set $\{\pm 1,..,\pm (L/2-1),L/2\}$. 

Employing Eq.~\eqref{number-fluc}, the particle number fluctuation, starting from an arbitrary initial state of the lattice, can be computed by solving Eq.~\eqref{eq:2pt_effective} and Eq.~\eqref{eq:4pt_effective}. The computation of the two-point correlator $D_{mm}$ involves matrices of size $O(L \times L)$, whereas for the four-point correlator $F_{mnmn}$ it is $O(L^2 \times L^2)$, which makes numerical simulation beyond lattice size $L \sim 10^2$ extremely difficult. For such small lattice sizes, we numerically find that it is not possible to observe non-equilibrium dynamics and extract clear scaling exponents without encountering finite-size effects. In particular, this persists for small values of the long-range exponent $\mu$. To circumvent this numerical challenge, we devise a novel bond-length representation that works suitably for any translationally invariant initial state, for example, alternating initial state. 

{\it Bond length representation for the four-point correlator.--} \df{We now present the} bond length representation starting from the effective equations of the four-point correlators, given in Eq.~\eqref{eq:4pt_effective}. \df{The advantage of this representation is twofold. First, using this approach, the computation of four-point correlators only involves a matrix of size $O(L \times L)$, which significantly reduces the numerical cost and enables the simulation of quantum dynamics for lattice sizes up to $\sim 10^4$. Second, it allows an analytical derivation of the particle number fluctuation as a function of time and and its FV scaling exponents in the short-range limit ($\mu \to \infty$).}
 
For a diagonal four-point correlator $F_{mnmn}$, we define its {\it bond length} as the distance $d(m,n)$ between the lattice sites $m$ and $n$. Under the periodic boundary condition, this distance is given as $d(m,n)= \min(|{m-n}|,L-|{m-n}|)$ and can take values $1, 2, \cdots L/2$. Interestingly, due to the translational symmetry of the lattice as well as that of the initial state, the four-point correlator $F_{mnmn}$ maintains this symmetry throughout its evolution. This implies that all $F_{mnmn}$ with the same bond length $p = d(m,n)$ exhibit the same dynamics. We denote all such elements with the single bond-length variable $G_p$. 
In this representation, Eq. (\ref{eq:4pt_effective}) reduces to the following equation of motion (see supplemental material \cite{supp} for the details) 
\begin{equation}
\label{eq:bond_length}
    \partial_t G_p = 2\,\Lambda\sum_{\substack{p' = 1 \\ p' \neq p}}^{L/2} \, U_{pp'}\left(G_{p'}-G_{p}\right),
\end{equation}
where the elements $U_{pp'}$ are given as follows. For $p'=\frac{L}{2}$,
\begin{equation}
U_{pp'} = \frac{1}{|{p-p'}|^{2\mu}}
\end{equation}
and for $p'<\frac{L}{2}$,
\begin{eqnarray}
    &U_{pp'} = \frac{1}{|{p-p'}|^{2\mu}} +
    \begin{cases}
            \frac{1}{(p+p')^{2\mu}} & \mathrm{if} \,\,  p+p' \leq \frac{L}{2} \\
            \frac{1}{(L-(p+p'))^{2\mu}} & \mathrm{if} \,\,  p+p'>\frac{L}{2}
        \end{cases}
\end{eqnarray}
As can be seen, computation of Eq.~\eqref{eq:bond_length} involves dealing only with $O(L \times L)$ matrix and thus, allows for accessing large systems. In what follows, we present results for the dynamics of bipartite particle number fluctuation, obtained using Eq.~\eqref{eq:2pt_effective} and Eq.~\eqref{eq:bond_length}.
\begin{figure}
    \centering
    \includegraphics[width=\linewidth]{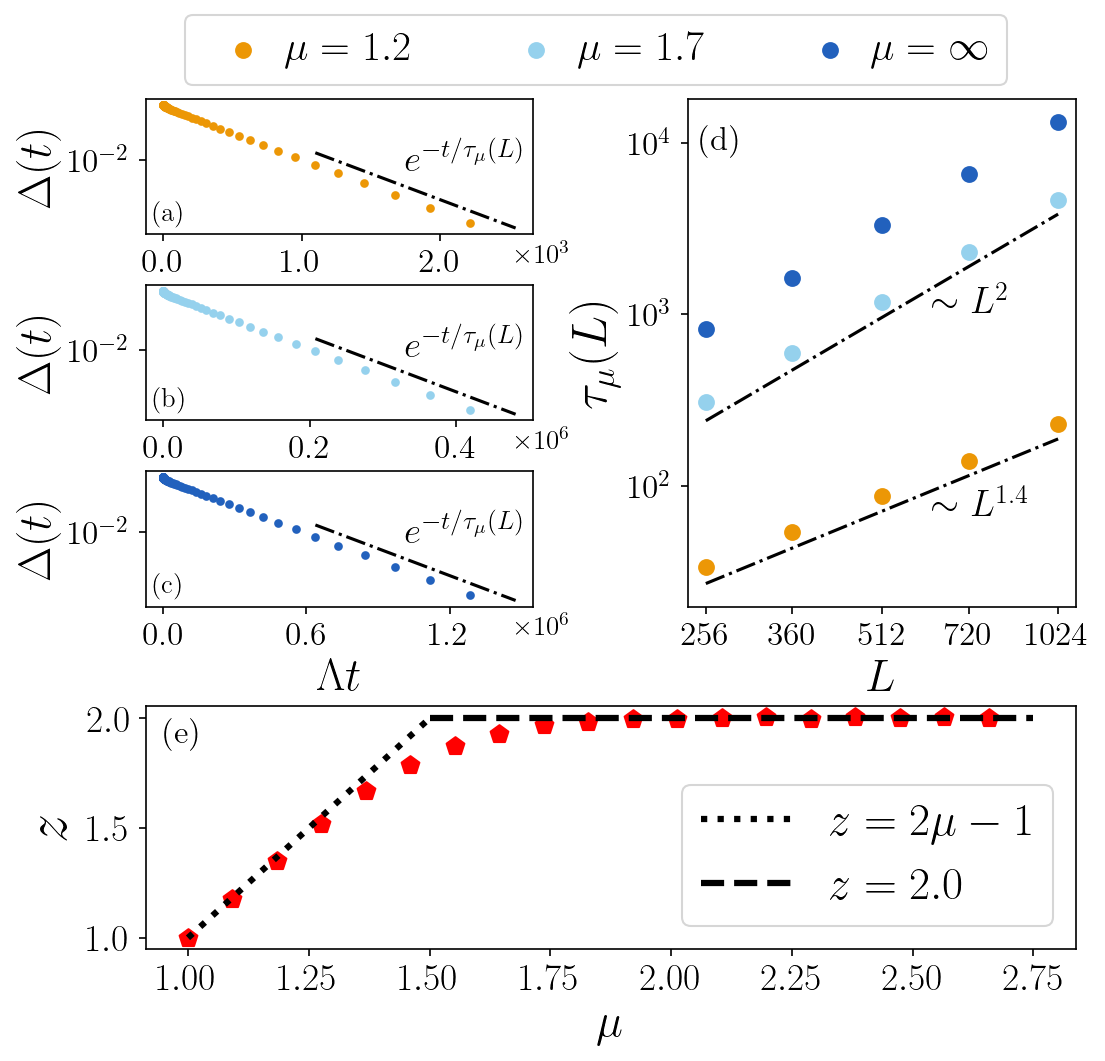}
    \caption{Plot for $\Delta(t)$ as defined in Eq.~\eqref{eq:deviation} to show the exponential relaxation of particle number fluctuation to the equilibrium value at late times (time when the value of $\Delta(t)$ reaches smaller than $10^{-2}$) for different values of long-range hopping. For (a) $\mu=1.2$, for (b) $\mu=1.7$, and for (c) $\mu \to \infty$, which corresponds to the tight-binding lattice. (d) Plot for the dependence of relaxation rate $\tau_{\mu}(L)$ on  lattice size $L$, $\tau_{\mu}(L)\sim L^z$ for different $\mu$ values. (e) Plot for the dynamical exponent $z$ v/s $\mu$  showing a crossover from superdiffusive ($z=2\mu-1$) to diffusive ($z=2.0$) regime at $\mu \sim 1.5$.}
    \label{fig:phase_diag}
\end{figure}

{\it Results.--} We first present results for two and four-point correlation functions for the long-range setup, starting with alternating initial state.
In Fig.~\ref{fig:dynamics_one_two}(a), we plot the dynamics of the diagonal elements of the two-point correlator, i.e.,  particle density at site $m$, $D_{mm}=\langle n_m \rangle$. We observe that the local occupation relaxes exponentially to the long-time value $1/2$ with a decay time scale given by $\tau_D=1/4\,\Lambda$, with $\Lambda=2J^2/\gamma$ being the effective diffusion constant. This behaviour can be understood by analytically solving Eq.~\eqref{eq:2pt_effective} in the Fourier representation, which gives, 
\begin{equation}
\label{eq:one_particle_soln}
D_{mm}(t) = \frac{1}{2}\left(1 + e^{i\pi m}\,e^{-E(k=L/2)t}\right),
\end{equation}
where 
\begin{equation}
    E(k) = \Lambda\left(4\sum_{j=1}^{\frac{L}{2}-1}\frac{\sin^2(\pi j k/L)}{j^{2\mu}} + \frac{1-e^{i \pi k}}{(L/2)^{2\mu}}\right)
    \label{eq:one_particle_dispersion}
\end{equation}
is the dispersion relation for the effective equations of the two-point correlator and $k=0,1, \cdots L-1$. It is easy to see that for the short-range hopping case ($\mu  \to \infty$), $E(k=L/2)=4\,\Lambda$. 
Thus, the local density relaxes exponentially to a maximally mixed state $D_{mm}(t \to \infty)=1/2$ with decay time $\tau_D= 1/4\Lambda$ and is independent of $L$. We also find that this result holds for long-range lattices as well (see Fig.~\ref{fig:dynamics_one_two}(a)).

Furthermore, following Eq.~\eqref{eq:one_particle_soln}, we find $\displaystyle \sum_{m=0}^{L/2-1}D_{mm}(t)=L/4$. That is, the number of particles in half the lattice system is a constant of motion. This implies that for the alternating initial state, the temporal evolution of particle number fluctuation is entirely governed by the four-point correlation function, given as $\displaystyle  \sigma^2(L,t) = -\sum_{m,n=0}^{L/2-1}F_{mnmn}(t) + \frac{L}{4} \big(1-\frac{L}{4}\big)$.  In Fig.~\ref{fig:dynamics_one_two}(b), we plot the dynamics of the diagonal elements of the four-point correlator, i.e., $F_{mnmn}$ and observe that, unlike the case for local density, the non-local four-point correlators saturate on a much longer and system-dependent time scale [see inset of Fig.~\ref{fig:dynamics_one_two}(b)]. The four-point correlator further exhibits interesting non-monotonic temporal evolution and eventually saturates to a value $\approx -1/4$. This saturation value can be predicted from the long-time limit of the density matrix, which settles to a maximally mixed state for which $F_{mnmn}(t\to \infty)=-(L-2)/(4L-4)$. For large $L$, this reduces to $F_{mnmn}=-1/4$. This result predicts the saturation value for the bipartite particle number fluctuation as $\displaystyle \lim_{t\to\infty} \sigma^2(L,t)\approx L/8$, independent of the value $\mu$. In Fig.~\ref{fig:dynamics_one_two}(c), we plot the dynamics for number fluctuation and observe diffusive growth characterized by scaling $\sigma^2(L,t) \sim t^{\beta}$ with $\beta=0.5$  for hopping exponent $\mu>1.5$ and super-diffusive growth $\sigma^2(L,t) \sim t^{\beta}$ with $\beta>0.5$ for $\mu<1.5$, followed by eventual saturation.
\begin{figure*}
    \centering
    \includegraphics[width=\linewidth]{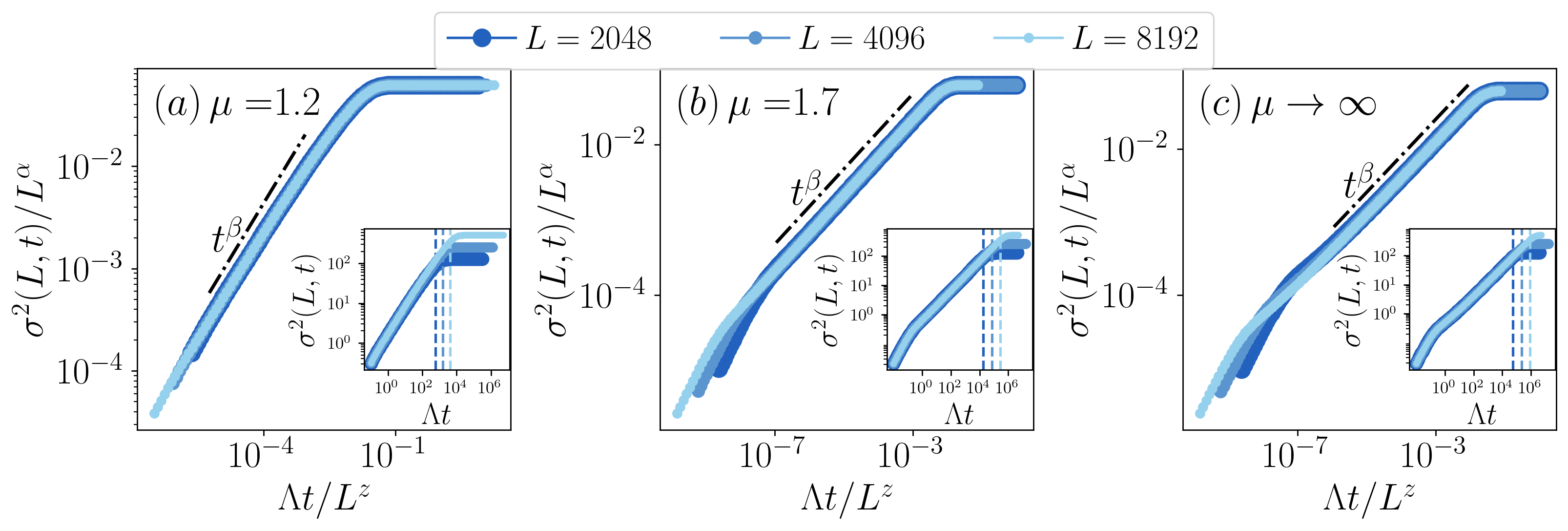}
    \caption{Plot for the growth and subsequent saturation of bipartite particle number fluctuation $\sigma^2(L,t)$ for a subsystem of size $L/2$ for the long-range hopping lattice model for different values of hopping exponent $\mu$ and different system sizes. The emergence of FV scaling [see Eq.~\eqref{eq:FV_scaling}] is evident once the ordinate and abscissa are normalized by $1/L^{\alpha}$ and $1/L^{z}$, respectively. We find super-diffusive exponents $(\alpha,\beta,z)= (1.0,1/(2\mu-1),2\mu-1$ for $\mu<1.5$ and diffusive scaling exponents $(\alpha,\beta,z)= (1.0,0.5,2.0) $ for $\mu \geq 1.5$. The inset of the figures shows unscaled dynamics for number fluctuation. The vertical dashed lines show the saturation time scale $\tau_{\mu}(L)$.}
    \label{fig:dynamics_fluc}
\end{figure*}
As expected from dissipative Lindblad dynamics [Eq.~\eqref{eq:master_eq}], $\sigma^2(L,t)$ relaxes exponentially towards its equilibrium value with a system-dependent time scale $\tau_{\mu}(L)$. Defining the deviation from equilibrium value $\Delta(t)$ as
\begin{equation}
\label{eq:deviation}
    \Delta(t) = \frac {\Big[\lim_{t\to\infty} \sigma^2(L,t)\Big] - \sigma^2(L,t)}{\lim_{t\to\infty} \sigma^2(L,t)}
\end{equation}
we demonstrate the exponential decay of $\Delta(t)$, i.e., $\Delta(t) \sim e^{-{t}/{\tau_{\mu}(L)}}$, for late times ($\Delta(t)<10^{-2}$) for different values of the hopping exponent $\mu$ in Fig.~\ref{fig:phase_diag}(a),(b) and (c). The system size dependence of the saturation timescale $\tau_{\mu}(L)$ is governed by the dynamical exponent $z$ as $\tau_{\mu}(L)\sim L^z$ which we show in Fig.~\ref{fig:phase_diag}(d) and extract $z$. 
In Fig.~\ref{fig:phase_diag}(e), we report the values of the dynamical exponent. For the dephased long-range setup, we find 
\begin{equation}
\label{eq:dynamical_exponent}
    z = \begin{cases}
        2\mu-1 &\mathrm{if}\,\,\mu < 1.5 \\
        2 &\mathrm{if}\,\,\mu\geq 1.5. \\
    \end{cases}
\end{equation}
which indicates super-diffusive non-equilibrium dynamics for $\mu <1.5$ and diffusive non-equilibrium dynamics for $\mu>1.5$ with crossover happening at $\mu \sim 1.5$.  In Fig \ref{fig:dynamics_fluc}(a)-(c), we plot the dynamics and eventual saturation of bipartite particle number fluctuations $\sigma^2(L,t)$ for three representative values of the hopping exponent $\mu=1.2$, $\mu=1.7$ and $\mu \to \infty$, respectively. For each case, the dynamics for different lattice sizes $L$ exhibit excellent data collapse upon rescaling $\sigma^2(L,t)$ and $t$ axes by $1/L^{\alpha}$ and $1/L^z$, respectively, with the saturation exponent $\alpha=1$ and the dynamical exponent $z$ given by Eq.~\eqref{eq:dynamical_exponent}. Additionally, at times $t\ll \tau_{\mu}(L)\sim L^z$, the fluctuations grow as $\sigma^2(L,t)\sim t^{\beta}$, with $\beta=\alpha/z$. These results confirm the Family-Vicsek (FV) scaling of bipartite particle number fluctuations, given by 
\begin{equation}\label{eq:FV_scaling}
    \sigma^2(L,t) = L^{\alpha}f\left(\frac{t}{L^z}\right),
\end{equation}
where, the scaling function $f(y)$ satisfies the limiting behaviour
\begin{equation*}
    f(y) \sim \begin{cases}
        y^{\beta} &\text{for}\qquad y \ll 1 \\
        1&\text{for}\qquad y \gg 1
    \end{cases}
\end{equation*}
with $\beta=\alpha/z$. This implies the scaling $\sigma^2(L,t)\sim t^{\beta}$ in the limit $t\ll L^z$. Thus, we observe FV scaling with super-diffusive exponents $(\alpha,\beta,z)= (1.0,1/(2\mu-1),2\mu-1$ for $\mu<1.5$ and diffusive scaling exponents  $(\alpha,\beta,z)= (1.0,0.5,2.0) $ for $\mu \geq 1.5$. In the supplemental material \cite{supp}, we report the emergence of FV scaling with the same scaling exponents also for the domain wall initial condition, highlighting the robustness of the universality class with respect to the choice of initial state. Furthermore, we find that the number fluctuation growth for a dephased long-range lattice setup consisting of bosons also shows FV scaling, with the same values for the scaling exponents and the same crossover at $\mu\sim 1.5$, as in the case of fermions (see supplementary material \cite{supp}).

\df{\textit{Analytical solution of particle fluctuations in $\mu \to \infty$ limit.--}} We now use the bond length representation to analytically obtain the diffusive FV scaling exponents in the short-range hopping limit ($\mu \to \infty) $. In this case, solving for the bond-length equation, given in Eq.~\eqref{eq:bond_length}, in the thermodynamic limit ($L\to\infty$), provides the two-particle spectrum 
$\lambda^{(k)} = - 4\Lambda \sin^2({\pi k}/{L})$, where, $k=0,1,\cdots,L/2-1$ (see the supplementary material \cite{supp}). Interestingly, this spectrum is identical to the single-particle spectrum given in Eq.~\eqref{eq:one_particle_dispersion} for $\mu \to \infty$.  Note that $\lambda^{(k)}=0$ only for $k=0$, which corresponds to the equilibrium mode for the four-point correlator and for all other eigenvalues, $\lambda^{(k_1)}>\lambda^{(k_2)}$ if $k_1<k_2$. The saturation time scale $\tau_{\infty}(L)$ is governed by the slowest decaying mode ($k=1$), which gives the dynamical exponent $z$. We obtain,  $\tau_{\infty}(L)=1/|\lambda^{(k=1)}|\sim L^2$ which gives $z=2$.

Furthermore, using the two-particle spectrum, it is possible to derive an exact analytical expression for the diagonals of the four-point correlation function. Considering $\Lambda t\gg 1$ and thermodynamic limit ($L\to \infty$), we get (see the supplementary material \cite{supp}) 
\begin{equation}
\label{eq:two_particle_profile}
    F_{mnmn}(t) \approx -\frac{1}{4} + \frac{1}{8\sqrt{4\pi\Lambda t}}e^{-\frac{|m-n|^2}{16\Lambda t}}, \quad (m\neq n),
\end{equation}
which has a Gaussian form and features translational symmetry. This suggests a diffusive growth of the bipartite particle number fluctuation. We find that in the limit $t\ll\tau_{\infty}(L)$, $\sigma^2(L,t)$ grows as, $\sigma^2(L,t)\approx \sqrt{t/4\pi}$ (see supplementary material \cite{supp}) implying the exponent value $\beta=1/2$. Previously, we confirmed that the saturation value of the fluctuation scales linearly with $L$ i.e., $
\displaystyle \lim_{t\to\infty}\sigma^2(L,t) \sim L$, corresponding to a spatial scaling exponent $\alpha=1$. Thus, we obtain the full set of scaling exponents for the dephased short-range hopping model as $(\alpha, \beta, z)=(1.0, 0.5, 2.0)$.

{\it Summary.--} We have investigated the non-equilibrium dynamics of bipartite particle number fluctuations in a dephased long-range fermionic lattice with power-law decaying hopping. We found that these fluctuations exhibit one-parameter scaling behaviour in the form of Family-Vicsek scaling with super-diffusive scaling exponents for $\mu < 1.5$ and diffusive scaling exponents for $\mu \geq 1.5$. Additionally, we have shown that, for an alternating initial state, the local density relaxes exponentially to its steady-state value, while the non-local four-point particle-number correlations saturate gradually in a system-size-dependent fashion. To facilitate both numerical and analytical calculations, we have introduced a novel bond length representation for translationally invariant initial states. This significantly reduces numerical complexity and enables analytical results for particle number fluctuations and scaling exponents. 
The on-site dephasing we have considered in this work mimics an effective inter-particle interaction arising purely due to the dissipative part and breaks the Wick's theorem -- yet our system of interest is still non-interacting. We believe an interacting system -- albeit challenging both analytically and numerically -- would significantly alter the relaxation dynamics of the non-local observables, thus, remains an open question.

\section*{Acknowledgments}
BKA acknowledges CRG Grant No. CRG/2023/003377 from Science and Engineering Research Board (SERB), Government of India. 
LT and BKA acknowledge the National Supercomputing Mission (NSM) for providing computing resources of ‘PARAM Brahma’ at IISER Pune, which is implemented by C-DAC and supported by the Ministry of Electronics and Information Technology (MeitY) and DST, Government of India. BKA thanks Katha Ganguly for insightful discussions related to this project and acknowledges the hospitality of the International Centre of Theoretical Sciences (ICTS), Bangalore, India, under the associateship program.

\bibliography{references}

\onecolumngrid

\setcounter{equation}{0}
\setcounter{figure}{0}
\setcounter{section}{0}
\renewcommand{\theequation}{S\arabic{equation}}
\renewcommand{\thefigure}{S\arabic{figure}}

\newpage

\begin{center}
{\textbf{\underline{Supplementary Material}}}
\end{center}

\section{Effective equations for two-point and four-point correlation functions}
\label{sec:supp_effective_eq}
In this section, we provide a detailed derivation of the effective equations for the two-point and four-point correlation functions in the strong dephasing limit $(\gamma \gg J)$, as given in Eq.~\eqref{eq:2pt_effective} and Eq.~\eqref{eq:4pt_effective} of the main text. 

The dynamics of the density matrix $\hat{\rho}(t)$ of a fermionic lattice system with $L$ sites which is subjected to particle number conserving dephasing at each site is modelled by the Lindblad master equation as 
\begin{equation}
\label{eq:master_eq_supp}
    \frac{d}{dt} \hat{\rho} = -i\Big[\hat{H},{\hat{\rho}(t)}\Big] + \gamma\,\sum_{j=0}^{L-1} \Big(\hat{n}_j \, \hat{\rho}\,\hat{n}_j -\frac{1}{2}\{\hat{n}_j,\hat{\rho}\} \Big),
\end{equation}
where $\hat{H}$ is the Hamiltonian of the lattice and given in Eq.~\eqref{hamiltonian}. $\hat{n}_i=c_i^{\dagger}c_i$ is the particle number operator at site $i$ and $\gamma$ is the dephasing strength. The two-point and four-point correlators are defined as $D_{mn}={\rm Tr} \big[\hat{c}^{\dagger}_m\hat{c}_n \hat{\rho}\big]$ and $F_{mnpq}={\rm Tr}\big[\hat{c}^{\dagger}_m \hat{c}^{\dagger}_n \hat{c}_p\hat{c}_q \hat{\rho}\big]$, respectively. In the following, we derive the effective equations of motions for the evolution of $D_{mn}$ and $F_{mnpq}$ by considering strong dephasing limit $\gamma\gg J$. 

\subsection{Two-point correlation function}\label{sec:supp_2pt_effective}
In this subsection, we derive the effective equation for the two-point correlation function. Following the Lindblad master equation in Eq.~(\ref{eq:master_eq_supp}), the equations of motion for the single particle density matrix $D_{mn}={\rm Tr} \big[\hat{c}^{\dagger}_m\hat{c}_n\hat{\rho}\big]$ is obtained as
\begin{equation}
\label{eq:2pt_micro}
\partial_t D_{mn} = iJ\sum_{j=\pm 1}^{\pm(L/2-1),L/2}\frac{D_{m(n+j)}-D_{(m+j)n}}{|j|^{\mu}}+ \gamma \, (\delta^m_n -1)D_{mn}.
\end{equation}
Here, $(m+j)\equiv (m+j)\mod L$ and $(n+j)\equiv (n+j)\mod L$. Note that even though the Lindblad equation in Eq.~(\ref{eq:master_eq_supp}) involves quartic operators, the equation of motion of the two-point correlation function is closed and does not involve higher point correlators \cite{Demler}. In fact, such a property holds also arbitrary $n$- point correlator for this setup. In order to derive the effective equation of motion in the large dephasing limit ($\gamma \gg J)$, we adiabatically eliminate the coherences from Eq.~\eqref{eq:2pt_micro}. We do so by
\begin{enumerate}
    \item Obtaining coherences $D_{mn} (m\neq n)$ in terms of populations $D_{mm}$.
    \item Using the relations obtained in 1, in the equation of motions of the populations $D_{mm}$.
\end{enumerate}
We first rescale the time with the slower variable 
$t \to \Tilde{t} = Jt$. In this time scale, the coherence elements evolve adiabatically, i.e.,  we have $\partial_{\Tilde{t}}D_{mn} \ll \frac{\gamma}{J} D_{mn}$. Hence, we can neglect the time derivative for the coherences i.e., the left hand side of Eq.~\eqref{eq:2pt_micro} for $m \neq n$. Furthermore, due to strong dephasing  $D_{mn}\ll D_{mm}$. As a result, following Eq.~\eqref{eq:2pt_micro} we obtain the coherence elements as 
\begin{equation}
    D_{mn} = i\frac{J}{\gamma}\Big(\frac{D_{mm}-D_{nn}}{d(m,n)^{\mu}}\Big).
    \label{Dmn}
\end{equation}
Here, $d(m,n)$ denotes the distance between lattice sites $m$ and $n$. Substituting Eq.~\eqref{Dmn} in Eq.~\eqref{eq:2pt_micro} for the populations $D_{mm}$, we obtain an effective equation for the populations (i.e., the diagonal elements of the single particle density matrix or the two-point correlation matrix) as  
\begin{equation}
    \partial_t D_{mm} = \frac{2J^2}{\gamma} \sum_{j=\pm 1}^{\pm(L/2-1),L/2}\frac{D_{(m+j)(m+j)} - D_{mm}}{|j|^{2\mu}}
    \label{eq:supp_2pt_effective}
\end{equation}
which is given in Eq.~\eqref{eq:2pt_effective} of the main text. At this point, it is important to address the range of the variable $j$ in the summation of Eq.~\eqref{eq:supp_2pt_effective} which crucially depends on the boundary conditions of the lattice system. Note that the variable $j$ denotes the distance between sites $(m+j)$ and $m$, and it takes values such that the index $(m+j)$ goes over all lattice sites $\{1,2,\cdots,L\}$ except the site $ m$ as $j \neq 0$. For a lattice system with open boundaries, the distance between lattice sites $m$ and $n$ is simply $|m-n|$. Thus, under open boundary conditions, $j$ takes values $\{1-m,2-m,\cdots,L-m\}$ except $m$ in Eq.~\eqref{eq:supp_2pt_effective}. However, for a lattice with periodic boundaries, the distance between sites $m$ and $n$ is $\min(|m-n|,L-|m-n|)$. And thus, for a periodic lattice with even number of sites, $j$ takes values $\{\pm 1,\pm 2,\cdots,\pm(L/2-1),L\}$ in Eq.~\eqref{eq:supp_2pt_effective}.

In Fig.~\ref{fig:supp_2pt_effective_alternating} we show the results obtained for two-point correlator computed using the exact equation of motion given in Eq.~\eqref{eq:2pt_micro} and the effective equation of motion given in Eq.~\eqref{eq:supp_2pt_effective}. The results show perfect match in the strong dephasing limit.  
\begin{figure}
    \centering
    \includegraphics[width=\linewidth]{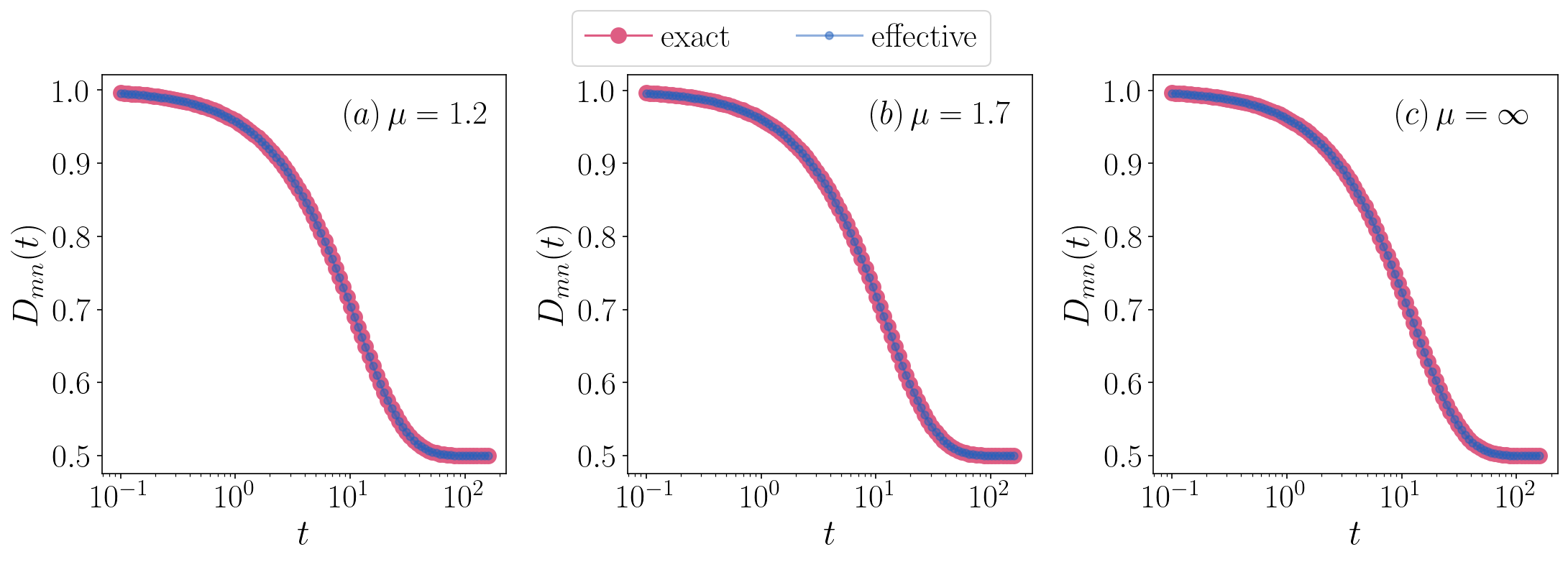}
    \caption{Plot for comparison of microscopic [Eq.~\eqref{eq:2pt_micro}] and effective equations [Eq.~\eqref{eq:supp_2pt_effective}] of the two-point correlator, under strong dephasing condition $(J=1,\gamma=100)$, for lattice size $L=32$. We plot here the local density $D_{mm}=\langle\hat{n}_{m}\rangle$ for site $m=0$, starting from alternating initial state.}
    \label{fig:supp_2pt_effective_V_exact}
\end{figure}

\subsection{Four-point correlation function}\label{sec:supp_4pt_effective}
In this subsection, we derive the effective equation for the four-point correlation function. Following the Lindblad master equation in Eq.~\eqref{eq:master_eq_supp}, the equation of motion for the two-particle density matrix $F_{mnpq}={\rm Tr}\big[\hat{c}^{\dagger}_m \hat{c}^{\dagger}_n \hat{c}_p\hat{c}_q \hat{\rho}\big]$  can be obtained as 
\begin{equation}
\label{eq:4pt_master_supp}
\begin{split}
\partial_t F_{mnpq} =& \, i\,J\sum_{j=\pm 1}^{\pm(L/2-1),L/2}\frac{(1-\delta^{p+j}_q)F_{mn(p+j)q}+(1-\delta^{p}_{q+j})F_{mnp(q+j)}-(1-\delta^{m+j}_{n})F_{(m+j)npq}-(1-\delta^{m}_{n+j})F_{m(n+j)pq}}{|j|^{\mu}}\\
&+\gamma\,(\delta_m^p + \delta_m^q + \delta_n^p + \delta_n^q - 2)F_{mnpq}.
\end{split}
\end{equation}
Here, $\delta^m_n$ denotes the Kronecker delta function, $(m+j)\equiv (m+j)\mod L$ and similarly for $(n+j)$,$(p+j)$ and $(q+j)$. The $(1-\delta^m_n)$ factors appearing in Eq.~\eqref{eq:4pt_master_supp} is due to the fermion statistics, which sets the correlators of the kind $F_{mnpp}$ and $F_{mmpq}$ to zero. Furthermore, fermionic anti-commutation relation for the creation and annihilation operators confers the following symmetry to the elements of the four-point correlation function, 
\begin{equation}
F_{mnpq}=-F_{mnqp}=-F_{nmpq}=F_{nmqp}.
\end{equation}
Hence there exist ${L\choose 2}^2$ linearly independent elements $F_{mnpq}$.  Out of these ${L \choose 2}^2$ linearly independent elements there are ${L \choose 2}$ diagonal elements which are of the form $F_{mnmn}$ or $F_{mnnm}$. On the other hand, we categorize the off-diagonal elements of $F_{mnpq}$ as follows, 
\begin{enumerate}
            \item Only one element of pair $(m,n)$ matches with pair $(p,q)$ i.e., either $m=p$, or $m=q$, or $n=p$, or  $n=q$.
            \item No element of pair $(m,n)$ matches with pair $(p,q)$. 
\end{enumerate}
Note that the equation of motion for the populations in Eq.~\eqref{eq:4pt_master_supp} connects the coherences of only type 1. In what follows, we now get an effective equation of motion that involves only the population elements i.e., $F_{mnmn}$ or $F_{mnnm}$ in the strong dephasing limit $\gamma \gg J$ by adiabatically eliminating the coherences. For this purpose, we first simplify the notation and denote the coherences of type 1 as $F_{c}$ and the populations as $F_{p}$. We once again introduce the slow timescale $\Tilde{t}=Jt$ and impose the strong dephasing limit where the coherences evolve adiabatically i.e.,  (i) $\partial_{\Tilde{t}}F_{c} \ll \frac{\gamma}{J} F_{c}$ and (ii) $F_{c}\ll F_{p}$. Applying these conditions in Eq.~\eqref{eq:4pt_master_supp} for $F_{c}$, gives us the coherence elements of type 1 in terms of the populations as, 
\begin{equation}
\label{eq:coher_to_population_supp}
\begin{split}
     F_{(m+j)nmn} &= (i{J}/{\gamma})(1-\delta^{m+j}_n)\Big({F_{(m+j)n(m+j)n}-F_{mnmn}}\Big)/{|j|^{\mu}}. \\
     F_{m(n+j)mn} &= (i{J}/{\gamma})(1-\delta^{m}_{n+j}) \Big({F_{m(n+j)m(n+j)}-F_{mnmn}}\Big)/{|j|^{\mu}}. \\
    F_{mn(m+j)n} &= (i{J}/{\gamma})(1-\delta^{m+j}_n)\Big({F_{mnmn}-F_{(m+j)n(m+j)n}}\Big)/{|j|^{\mu}}.\\
    F_{mnm(n+j)} &= (i{J}/{\gamma})(1-\delta^m_{n+j})\Big({F_{mnmn}-F_{m(n+j)m(n+j)}}\Big)/{|j|^{\mu}}.
\end{split}
\end{equation}
Here again, the $(1-\delta^m_n)$ factor appear due to the fermionic statistics which implies $F_{mmmn}=F_{mnmm}=0$ which are obeyed by the above solutions. Substituting for these coherence elements given in Eq.~\eqref{eq:coher_to_population_supp} in the equation of motion for the populations $F_{mnmn}$ with $(m\neq n)$, gives the effective equations for four-point correlation function elements as
\begin{equation}
\label{eq:supp_4pt_effective}
    \partial_t F_{mnmn} = \frac{2J^2}{\gamma} \sum_{j=\pm 1}^{\pm(L/2-1),L/2}\frac{F_{m(n+j)m(n+j)}+F_{(m+j)n(m+j)n}-2F_{mnmn}}{|j|^{2\mu}}
\end{equation}
which is given in Eq. (\ref{eq:4pt_effective}) of the main text.
\begin{figure}
    \centering
    \includegraphics[width=\linewidth]{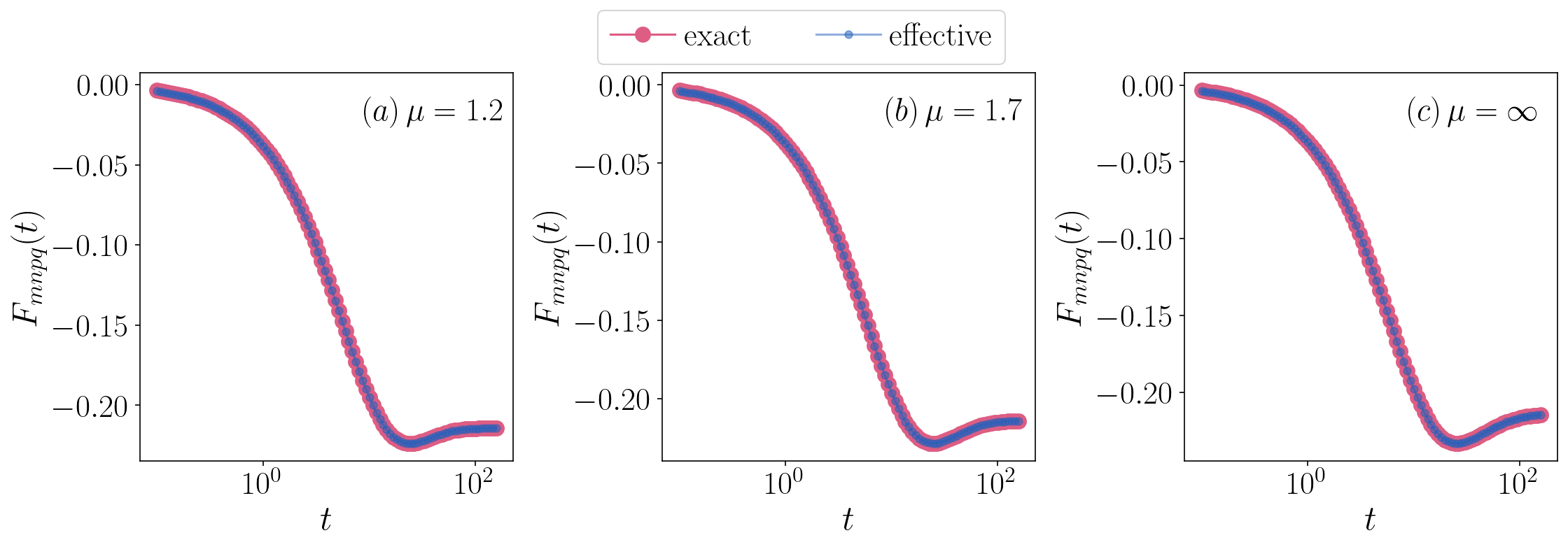}
    \caption{Plot for comparison of microscopic [Eq.~\eqref{eq:4pt_master_supp}] and effective equations [Eq.~\eqref{eq:supp_4pt_effective}] of the four-point correlator under strong dephasing condition $(J=1,\gamma=100)$, for lattice size $L=8$. We plot for the four-point correlator $F_{mnpq}=\langle\hat{c}^{\dagger}_m \hat{c}^{\dagger}_n \hat{c}_p\hat{c}_q\rangle$ for $(m,n,p,q)=(1,L/2,1,L/2)$, starting from alternating initial state.}
    \label{fig:supp_4pt_effective_V_exact}
\end{figure}
In Fig.~\ref{fig:supp_4pt_effective_V_exact} we show the results obtained for four-point correlator computed using the exact equation of motion given in Eq.~\eqref{eq:4pt_master_supp} and the effective equation of motion given in Eq.~\eqref{eq:supp_4pt_effective}. The results show perfect match in the strong dephasing limit. 

\section{Solution for effective two-point correlation function}\label{sec:supp_2pt_effective_soln}
In this section, we provide the solution for the effective equation for the two-point correlation function starting from alternating initial state. We further compare and contrast the relaxation time scales of the two-point correlation function, when the initial state is considered to be a domain wall state. Such a state represent large-scale density variation as compared to alternating initial state.  

The effective equation for the one-particle density matrix is given as:
\begin{equation}
    \partial_t D_{mm} = \Lambda \sum_{j=\pm 1}^{\pm(L/2-1),L/2}\frac{D_{(m+j)(m+j)} - D_{mm}}{|j|^{2\mu}},
    \label{eq:supp_2pt_effective-extra}
\end{equation}
where we denote $\Lambda=2J^2/\gamma$ as the diffusion constant. Under periodic boundary conditions, $j$ goes over the values $\{\pm 1,\pm 2,\cdot,\pm(L/2-1),L/2\}$ and the equation is translationally invariant. We define the discrete Fourier transform as
\begin{equation}
    \tilde{D}_k = \frac{1}{\sqrt{L}}\sum_{m=1}^{L}e^{-\frac{2\pi i}{L}mk} \, D_{mm},  \quad k\in\{0,1,\cdots,L-1\},
    \label{Fourier_k}
\end{equation}
and express Eq.~\eqref{eq:supp_2pt_effective-extra} in the Fourier domain as 
\begin{equation}
    \partial_t \tilde{D}_k = -E(k) \Tilde{D}_k,
    \label{linear-t}
\end{equation}
where the dispersion relation is given as 
\begin{equation}
    E(k) = \Lambda\left(4\sum_{j=1}^{\frac{L}{2}-1}\frac{\sin^2(\pi j k/L)}{j^{2\mu}} + \frac{1-e^{i \pi k}}{(L/2)^{2\mu}}\right).
    \label{eq:supp_one_particle_dispersion}
\end{equation}
In the long-time limit ($t\to\infty$), only the $k=0$ mode survives, since $E(k=0)=0$ and $E(k)>0$ for $k>0$. 

\begin{figure}
    \centering
    \includegraphics[width=\linewidth]{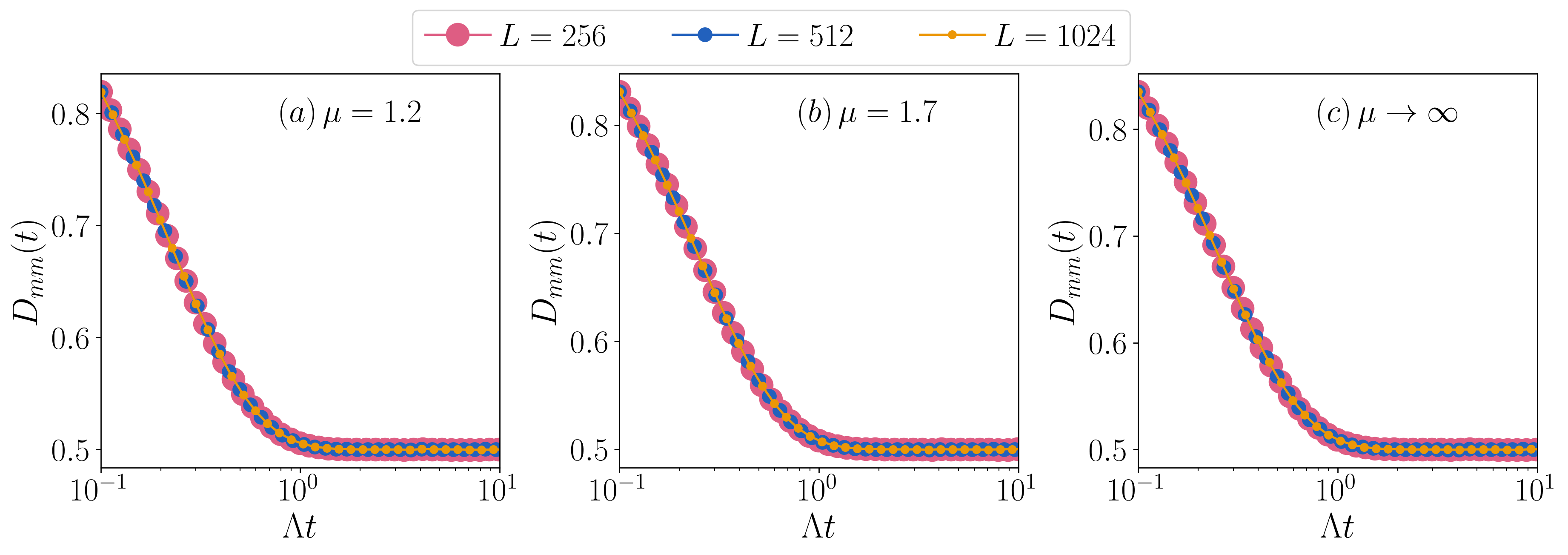}
    \caption{Plot for the diagonal elements of two-point correlator $D_{mm}$ with $m=0$, as a function of $t$ starting from alternating initial state for different values of power law hopping exponent $\mu$. For (a) $\mu=1.2$, (b) $\mu=1.7$, and (c) $\mu=\infty$, representing the short-range or tight-binding lattice, and different lattice sizes $L=256$,$L=512$ and $L=1024$. For all values of $\mu$, the two-point correlator relaxes in system size independent time-scale $\sim 1/\Lambda$.}
    \label{fig:supp_2pt_effective_alternating}
\end{figure}

{\it Alternating initial state.--} Now, for the alternating initial state for the lattice, i.e., $|\psi(0)\rangle = \prod_{m=0}^{L/2-1}c^{\dagger}_{2m} |0 \rangle$  we have $D_{mm}(t=0)=1$ for even $m$ and $D_{mm}(t=0)=0$ for odd $m$. For this initial state, $\tilde{D}_k$ at the initial time can be expressed as 
\begin{equation}
\begin{split}
    \Tilde{D}_k(t=0) &= \frac{1}{\sqrt{L}}\sum_{m=1}^{L}e^{-\frac{2\pi i}{L}mk} D_{mm}(t=0) \\
    &= \frac{1}{\sqrt{L}}\sum_{m=1}^{L/2}e^{-\frac{2\pi i}{L/2}mk} \\
    &= \frac{\sqrt{L}}{2}\left(\delta^k_0 + \delta^k_{L/2} \right),
\end{split}
\end{equation}
where we assume $L$ to be even. This implies that at $t=0$ only two Fourier modes $k=0$ and $k=L/2$ are occupied. Therefore, following Eq.~\eqref{Fourier_k} and Eq.~\eqref{linear-t}, $D_{mm}(t)$ is given as
\begin{equation}
\begin{split}
    D_{mm}(t) &= \frac{1}{\sqrt{L}}\sum_{m=1}^{L}e^{\frac{2\pi i}{L}mk} \, \Tilde{D}_k(t) \\
    &= \frac{1}{\sqrt{L}}\left(\Tilde{D}_{k=0}(t=0) + e^{i\pi m}\, e^{-E(k=L/2)t}\,\Tilde{D}_{k=L/2}(t=0) \right)\\
    &= \frac{1}{2}\left(1 + e^{i\pi m}\,e^{-E(k=L/2)t}\right).
\end{split}
\end{equation}
Now, for the tight-binding lattice case, i.e., for nearest-neighbour hopping ($\mu \to \infty$) we get from Eq.~\eqref{eq:supp_one_particle_dispersion} $E(k=L/2)=4\,\Lambda$. Hence, for the alternating initial state, the one-particle density matrix for the tight-binding lattice exponentially relax to a maximally mixed state $D_{mm}(t \to \infty)=1/2$ with relaxation time scale given as $1/(4\Lambda)$, which is governed by the diffusion constant $\Lambda=2J^2/\gamma$ and is independent of the lattice size $L$. In Fig.~\eqref{fig:supp_2pt_effective_alternating}(c), we plot the behaviour of $D_{mm}(t)$ for alternating initial state for the tight-binding case and find exact match with our analytical results. We further find that, this relaxation time scale holds for long-range lattice as well for arbitrary values of $\mu$. We show this in Fig.~\eqref{fig:supp_2pt_effective_alternating}(a) and (b) for two different values of $\mu$, $\mu=1.2$, and $\mu=1.7$, respectively. 

\begin{figure}
    \centering
    \includegraphics[width=\linewidth]{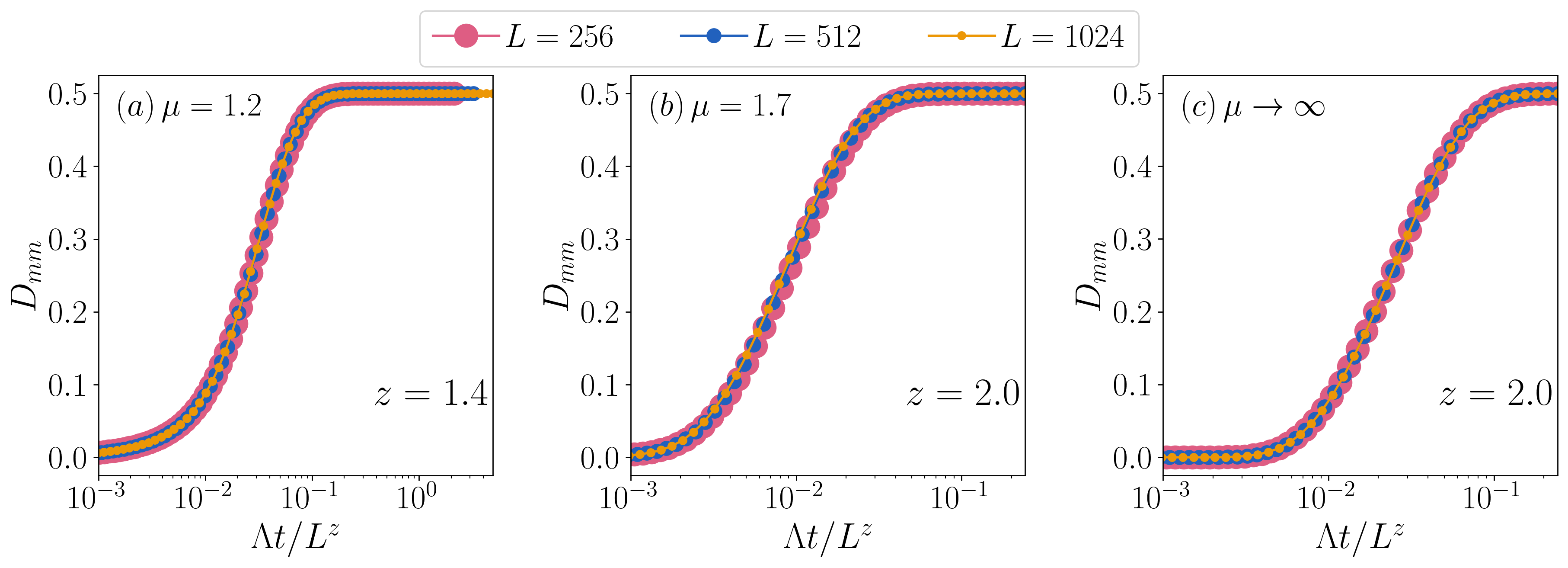}
    \caption{Plot for the diagonal elements of two-point correlator $D_{mm}$ with $m=3L/4$, as a function of $t$ starting from the domain wall initial state for different values of power law hopping exponent $\mu$. For (a) $\mu=1.2$, (b)$\mu=1.7$, and (c)$\mu \to \infty$, representing the short-range lattice, and different lattice sizes $L=256$,$L=512$ and $L=1024$. For the domain wall initial state, saturation time scale of the two-point correlator shows system size scaling dependence, as illustrated by the collapse in data upon re-scaling the $x$-axis by $1/L^z$. For $\mu<1.5$, we have $z=2\mu-1$, whilst for $\mu\geq1.5$ we have $z=2.0$.}
    \label{fig:supp_2pt_effective_domain_wall}
\end{figure}

{\it Domain wall initial state.--} 
Interestingly, we see a completely different relaxation dynamics for the domain wall initial state which shows a system-size dependent relaxation time scale.  Considering the left half of the lattice sites to be initially occupied, i.e., $|\psi(0)\rangle = \prod_{m=1}^{L/2}c^{\dagger}_{m} |0 \rangle$, $D_{mm}(t=0)=1$ for $m=1,2, \cdots L/2$ and $D_{mm}(t=0)=0$ for $m=L/2+1, \cdots L$. The inverse Fourier transform of $D_{mm}(t=0)$ shows that $k=0$ mode and all the odd $k$ modes are occupied at $t=0$:
\begin{equation}
\begin{split}
    \Tilde{D}_k(t=0) &= \frac{1}{\sqrt{L}}\sum_{m=1}^{L}e^{-\frac{2\pi i}{L}mk} D_{mm}(t=0) \\
    &= \frac{1}{\sqrt{L}}\sum_{m=1}^{L/2}e^{-\frac{2\pi i}{L}mk}
\end{split}  
\end{equation}
The summation evaluates to:
\begin{equation}
    \sum_{m=1}^{L/2}e^{-\frac{2\pi i}{L}mk} \begin{cases}
        = L/2 & \;\text{for } k=0 \\
        \begin{cases}
            = 0 &\,\text{for } k \in \mathrm{even} \\
            \neq 0 &\,\text{for } k \in \mathrm{odd}
        \end{cases} & \;\text{for } k>0  
    \end{cases}
\end{equation}
The lowest lying non-zero mode $(k=1)$ decays the slowest and determines the time scale of equilibration. Hence, in the long time limit $(t\to\infty)$, we may neglect modes $k>1$:
\begin{equation}
\begin{split}
    D_{mm}(t) &= \frac{1}{\sqrt{L}}\sum_{m=1}^{L}e^{\frac{2\pi i}{L}mk} \, \Tilde{D}_k(t) \\
    & \approx \frac{1}{\sqrt{L}}\left(\Tilde{D}_{k=0}(t=0) + e^{\frac{2\pi i}{L}m}\, e^{-E(k=1)t}\,\Tilde{D}_{k=1}(t=0) \right).
\end{split}
\end{equation}
where in the second line we have considered long-time limit. 
For tight binding lattice $(\mu \to \infty)$, $E(k=1)\approx 4\pi^2\Lambda/L^2$. Hence, starting from the domain wall state, the diagonals of two-point correlation function, i.e., the local density at each site relaxes to equilibrium with time scale $t^*\sim L^2$. This we illustrate in Fig.~\ref{fig:supp_2pt_effective_domain_wall}(c). We numerically find the system size scaling of this relaxation time scale $t^*$ for long-range lattice is given by 
\begin{equation}
    t^*=\frac{1}{E(k=1)}\sim\begin{cases}
        L^{2\mu-1} &\text{for}\;\mu<1.5\\
        L^2 &\text{for}\;\mu\geq1.5
    \end{cases}
\end{equation}
We illustrate this system size scaling in Fig.~\ref{fig:supp_2pt_effective_domain_wall}(a),(b) for two different values of $\mu$, $\mu=1.2$ and $\mu=1.7$, respectively.

\section{Bond length representation for four-point correlation function}
\label{sec:supp_bond_length_derivation}
In this section, we put forward a suitable representation, namely the {\it bond-length} representation, for solving the effective four-point correlation function, given in Eq.~\eqref{eq:supp_4pt_effective}, for alternating initial state. 
As mentioned in the main text, lattice consisting of $L$ sites, solving for the four-point correlator $F_{mnmn}$ requires dealing  with matrix of size $O(L^2 \times L^2)$. This makes numerical simulation beyond lattice size $L \sim 10^2$ extremely challenging. In fact, for such small lattice sizes, we numerically find that it is not possible to observe non-equilibrium dynamics and extract clear scaling exponents without encountering finite-size effects. 
Therefore, there is a need to find an alternate way to simulate two-particle density matrix efficiently. We show that, for alternate initial condition, it is possible to find a so-called bond length representation where the one needs to deal with $O(L \times L)$ matrix and thus one achieves a significant numerical advantage. Furthermore, this representation is amenable to analytical calculations for particle number fluctuations. We now provide the details of this representation. 

For the diagonal elements of the two-particle density matrix, i.e., $F_{mnmn}=\tr[\rho\hat{c}^{\dagger}_m \hat{c}^{\dagger}_n \hat{c}_m\hat{c}_n]$, we define the distance between its indices, denoted by $d(m,n)$, as its bond length. Under the periodic boundary condition, we then have $d(m,n)=\min(|m-n|,L-|m-n|)$.
Now, for the alternating initial state, all bonds with $d$ being odd have one site occupied and another unoccupied. On the other hand, for bonds with $d$ being even, either both sites $m$ and $n$ are occupied, or both are unoccupied. Taking these points into consideration along with the periodicity of the lattice and that of the initial state, we make the following claims:

\begin{itemize}
    \item[--] Given a value $p\in\{1,3,..,L/2-1\}$, all diagonals $F_{mnmn}$ with bond length $d(m,n)=p$ are identical, and we label these as $G^{(0)}_p$. 
    That is, $F_{0101}=F_{1212}= \cdots = F_{(L-1)0(L-1)0}=G_{1}^{{(0)}}$, $F_{0303}=F_{1414}= \cdots = F_{(L-1)2(L-1)2}=G_{3}^{{(0)}}$ and so on for $p=5,7,\cdots,L/2-1$.
    \item[--] Given a value $p\in\{2,4,\cdots,L/2\}$, all diagonals $F_{mnmn}$ with (i) bond length $d(m,n)=p$ and (ii) both sites occupied in the initial state, are identical and we label these as $G^{(1)}_p$.
    That is, $F_{0202}=F_{2424}=\cdots=F_{(L-2)0(L-2)0}=G^{(1)}_2$, $F_{0404}=F_{2626}=\cdots=F_{(L-2)2(L-2)2}=G^{(1)}_4$ and so on for $p=6,8,\cdots,L/2$. 
    \item[--] Given a value $p\in\{2,4,\cdots,L/2\}$, all diagonals $F_{mnmn}$ with bond length $d(m,n)=p$ and both sites unoccupied in the initial state are identical, and we label these as $G^{(2)}_p$.
    That is, $F_{1313}=F_{3535}=\cdots=F_{(L-1)1(L-1)1}=G^{(2)}_2$, $F_{1515}=F_{3737}=F_{(L-3)1(L-3)1}=G^{(2)}_4$ and so on for $p=6,8,\cdots,L/2$.
\end{itemize}
The above rules define a bijection between the sets $\{F_{mnmn}\}$ and $\{G^{(0),(1),(2)}_p\}$. Applying this mapping to (\ref{eq:supp_4pt_effective}) gives:
\begin{equation}
\begin{split}
    \partial_t G^{(0)}_p = & 2\,\Lambda\,\sum_{p'\neq p,p'\in odd}U_{pp'}\big(G^{(0)}_{p'}-G^{(0)}_p \big) + 2\,\Lambda\,\sum_{p'\neq p,p'\in even}U_{pp'}\Big(\frac{G^{(1)}_{p'} + G^{(2)}_{p'}}{2}-G^{(0)}_p\Big) \\
    \partial_t G^{{(1)},{(2)}}_p = & 2\,\Lambda\sum_{p'\neq p,p'\in odd}U_{pp'} \big(G^{(0)}_{p'}-G^{{(1)},{(2)}}_p\big) +  2\Lambda\sum_{p'\neq p,p'\in even}U_{pp'}\big(G^{{(1)},{(2)}}_{p'}-G^{{(1)},{(2)}}_p\big)
    \label{four-particle}
\end{split}
\end{equation}
where recall that $\Lambda=2J^2/\gamma$ is the effective diffusion constant. The coupling matrix elements $U_{pp'}$ in Eq.~\eqref{four-particle} are given by
\begin{equation}
    U_{pp'} = \begin{cases}
        \frac{1}{|p'-p|^{2\mu}} & \text{if  $p'=L/2$} \\
        \frac{1}{|p'-p|^{2\mu}} + 
        \begin{cases}
            \frac{1}{(p+p')^{2\mu}} & \text{if  $p+p' \leq L/2$} \\
            \frac{1}{(L-(p+p'))^{2\mu}} & \text{if  $p+p'>L/2$}
        \end{cases} & \text{if $p'<L/2$}
    \end{cases}
\end{equation}
Note that for $p\in even$, the occupied bonds i.e., $G^{(1)}_p$ and the unoccupied bonds i.e., $G^{(2)}_p$ contribute equally in Eq.~\eqref{four-particle}. Therefore instead of working with these two quantities separately, we define a single quantity as an arithmetic mean of the two variable $(G^{(1)}_p + G^{(2)}_p)/2$, for all $p \in even$. In other words for $p \in even$, we use a representation where we average over the occupied and unoccupied bonds:
\begin{equation}
    G_p = \begin{cases}
        G^{(0)}_p & \text{for } p\in \mathrm{odd} \\
        (G^{(1)}_p + G^{(2)}_p)/2 & \text{for } p \in \mathrm{even}
    \end{cases}
\end{equation}
Finally, we receive the equations of motion in terms of the bond variables as,
\begin{equation}
\label{eq:bond_lenght_supp}
    \partial_t G_p = 2\,\Lambda\,\sum_{p'\neq p}U_{pp'}\, (G_{p'}-G_p)
\end{equation}
which matches with Eq.~\eqref{eq:bond_length} of the main text.
\begin{figure}
    \centering
    \includegraphics[width=\linewidth]{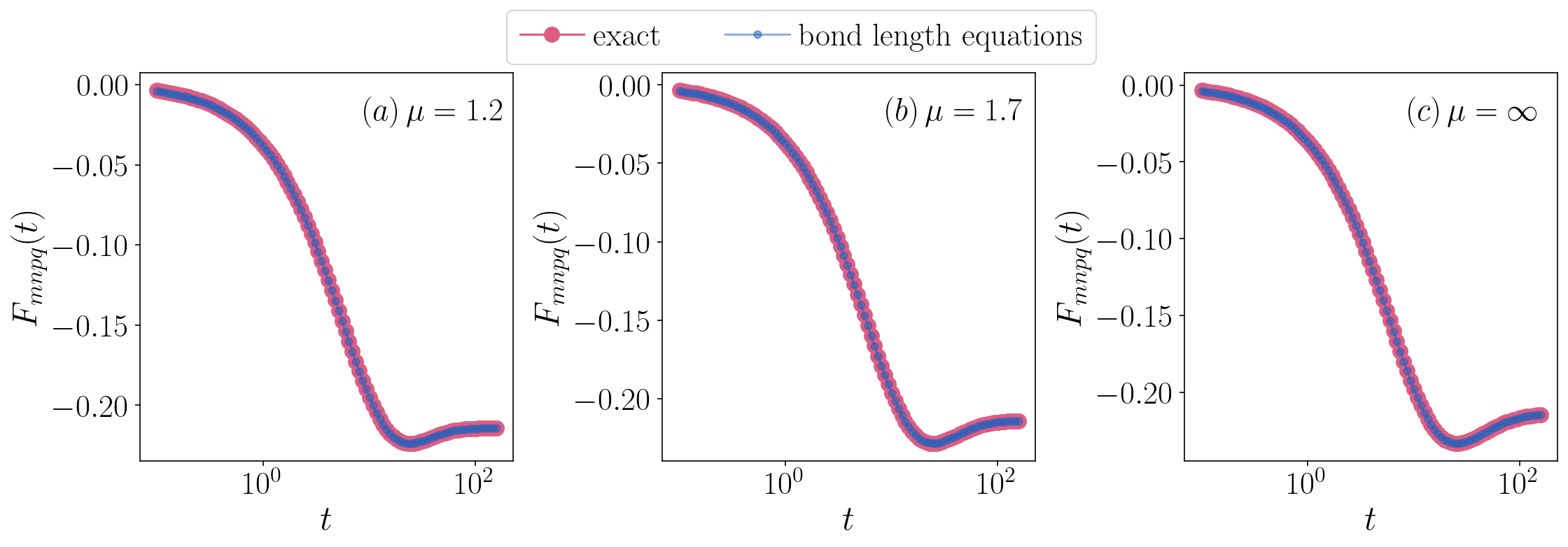}
    \caption{Plot for comparison of microscopic (Eq.~\eqref{eq:4pt_master_supp}) and bond length equations (Eq.~\eqref{apndx_eq:4pt_effective}) of the two-point correlator, under strong dephasing conditions $(J=1,\gamma=100)$, for lattice size $L=8$. $x$-axis plots for the two-point correlator $F_{mnpq}=\langle\hat{c}^{\dagger}_m \hat{c}^{\dagger}_n \hat{c}_p\hat{c}_q\rangle$ for $(m,n,p,q)=(1,L/2,1,L/2)$, starting from the alternating initial state.}
    \label{fig:supp_4pt_bond_len_V_exact}
\end{figure}

\section{Exact analytical solution for four-point correlation function in the bond length representation for tight-binding lattice ($\mu \to \infty$)}
In this section, we analytically solve the bond length representation of the effective equations of the two-particle density matrix, starting from alternating initial state, $\hat{\rho}_{\text{alt}}=|\psi_{\text{alt}}\rangle\langle \psi_{\text{alt}}|$ with $|\psi_{\text{alt}}\rangle=\prod_{m=0}^{L/2-1}\hat{c}^{\dagger}_{2m} |0 \rangle$, in the thermodynamic limit of the lattice size $L$.

The bond length equations are derived in Sec.~\eqref{sec:supp_bond_length_derivation}. We solve the equations for the case of tight-binding Hamiltonian, i.e., the hopping exponent $\mu$ is set to $\infty$ in the bond length equations Eq.~\eqref{eq:bond_lenght_supp}. We begin by writing Eq.~\eqref{eq:bond_lenght_supp} as a linear, first-order ordinary differential equation with a rescaling in time $t \to \tau=2\Lambda t$ which gives 
\begin{equation}
    \frac{\partial}{\partial \tau}|G\rangle = \mathbf{U}_{\mu}|G\rangle.
\end{equation}
Where, $p$-th element of the vector $|G\rangle$ is the bond length variable $G_p$. The matrix $\mathbf{U}_{\mu}$ here is of dimension $L/2$. For the tight binding case ($\mu \to \infty$) the generator of the dynamics $\mathbf{U}_{\mu \to \infty}$ is tridiagonal and given as
\begin{equation}
    \mathbf{U}_{\mu \to \infty} = \begin{pmatrix}
                    -1 & 1 & 0 & 0 & \ldots & 0 & 0 \\
                    1 & -2 & 1 & 0 & \ldots & 0 & 0 \\
                    0 & 1 & -2 & 1 & \ldots & 0 & 0 \\
                    \ldots & \ldots & \ldots & \ldots & \ldots & \ldots & \ldots \\
                    0 & 0 & \ldots & 1 & -2 & 1 & 0 \\
                    0 & 0 & 0 & \ldots & 1 & -2 & 1 \\
                    0 & 0 & 0 & \ldots & 0 & 2 & -2 \\
                    
                    \end{pmatrix}_{L/2 \times L/2}.
\end{equation}
The generator $\mathbf{U}_{\mu}$ is not fully symmetric due to the presence of the last row. However it can be symmetrized by re-scaling the last row by half. We denote the symmetrized generator as $\Tilde{\mathbf{U}_{\mu}}$. We define 
\begin{equation}
    S = \begin{pmatrix}
        1 & 0 & 0 & \ldots & 0 \\
        0 & 1 & 0 & \ldots & 0 \\
        \ldots & \ldots & \ldots & \ldots & \ldots \\
        0 & \ldots & 0 & 1 & 0 \\
        0 & \ldots & 0 & 0 & \frac{1}{2} \\
    \end{pmatrix}_{L/2 \times L/2} \quad \mathrm{and}\quad 
    \Tilde{\mathbf{U}}_{\mu\to\infty} = S\mathbf{U}_{\mu\to\infty} = \begin{pmatrix}
                    -1 & 1 & 0 & 0 & \ldots & 0 & 0 \\
                    1 & -2 & 1 & 0 & \ldots & 0 & 0 \\
                    0 & 1 & -2 & 1 & \ldots & 0 & 0 \\
                    \ldots & \ldots & \ldots & \ldots & \ldots & \ldots & \ldots \\
                    0 & 0 & \ldots & 1 & -2 & 1 & 0 \\
                    0 & 0 & 0 & \ldots & 1 & -2 & 1 \\
                    0 & 0 & 0 & \ldots & 0 & 1 & -1 \\
                    
                    \end{pmatrix}_{L/2 \times L/2}
\end{equation}
We first show that the dynamics generated by $\mathbf{U}_{\mu\to\infty}$ and $\mathbf{\tilde{U}}_{\mu\to\infty}$ for a bond length variable $G_p(t)$, converges in the limit $L\to\infty$, provided that $p$ is finite i.e., $p \ll L$. For this, it suffices to show that in the limit $L\to\infty$, the $p$-th row of the propagators $e^{\mathbf{U}_{\mu\to\infty}t}$ and $e^{\mathbf{\tilde{U}}_{\mu\to\infty}t}$ are identical, which we prove as follows. Note that the propagators can be approximated to arbitrary precision, by choosing an arbitrary high upper cut-off $m$ in their series expansion,
\begin{equation}
\label{eq:supp_prop_expan}
    e^{\mathbf{U}_{\mu\to\infty}t} \approx \sum_{n=0}^{m}\frac{1}{n!}\mathbf{U}^n_{\mu\to\infty}t^n \qquad \qquad e^{\mathbf{\tilde{U}}_{\mu\to\infty}t}\approx \sum_{n=0}^{m}\frac{1}{n!}\mathbf{\tilde{U}}^n_{\mu\to\infty}t^n
\end{equation}
Further, the tridiagonal structure of the generators $\mathbf{U}_{\mu\to\infty}$ and $\mathbf{\tilde{U}}_{\mu\to\infty}$ implies that the matrices $\mathbf{U}^n_{\mu\to\infty}$ and $\mathbf{\tilde{U}}^n_{\mu\to\infty}$ differ only from rows $\frac{L}{2}-n$ to $\frac{L}{2}$. Hence in the limit $L\to\infty$, one may choose an arbitrarily high cut-off $m$ in expansion Eq.~\eqref{eq:supp_prop_expan}, keeping the $p$-th row of the propagators $e^{\mathbf{U}_{\mu\to\infty}t}$ and $e^{\mathbf{\tilde{U}}_{\mu\to\infty}t}$ identical. Henceforth, we use $\mathbf{\tilde{U}}_{\mu\to\infty}$ to analyse the time dynamics of four-point correlator. The eigenvalues $\lambda_k$ and corresponding eigenvectors $|v^{(k)}\rangle$ of $\Tilde{\mathbf{U}}_{\mu\to\infty}$ are given as \cite{yueh2005eigenvalues}
\begin{equation}
\begin{split}
\label{eq:eigenpairs}
    \lambda_k &= -2 + 2\cos\bigg(\frac{2\pi k}{L}\bigg) = -4\sin^2\bigg(\frac{\pi k}{L}\bigg),\\
    |v^{(k)}\rangle_j &= \begin{cases}
        1/\sqrt{L/2}, & \text{for } k=0 \\
        \frac{1}{\sqrt{L/4}}\cos\bigg(\frac{\pi k(2j-1)}{L}\bigg),  & \text{for }    k\in\{1,..,L/2-1\}
    \end{cases} \qquad 
\end{split}
\end{equation}
where $j=1,\ldots,L/2$. The bond length vector $|G(\tau)\rangle$ at time $\tau=2\Lambda t$ is given by
\begin{equation}
    |G(\tau)\rangle = \mathcal{F}^{-1}e^{\mathcal{F}\Tilde{U}\mathcal{F}^{-1}\tau}|\Tilde{G}(0)\rangle,
\end{equation}
where $|\Tilde{G}(0)\rangle=\mathcal{F}|G(0)\rangle$, $k$-th row of $\mathcal{F}$ is the eigenvector $|{v}^{(k)}\rangle$ and $\mathcal{F}^{-1}=\mathcal{F}^{T}$. The alternating initial condition for the lattice in terms of bond length variables is given as as $|G(0)\rangle_p =0$ for $p=1,3,\cdots,L/2-1$ and $|G(0)\rangle_p =-\frac{1}{2}$ for $p=2,4,\cdots,L/2$. This state in the eigenbasis of $\tilde{\mathbf{U}}$ representation can be expressed as 
\begin{equation}
\begin{split}
 \Tilde{G}_k(\tau=0) &= \sum_{j=1}^{L/2} |v^{(k)}\rangle_j |G(0)\rangle_j \\
    &= \begin{cases}
        - \frac{L/8}{\sqrt{L/2}} & \text{for  } k=0 \\
        -\frac{1}{\sqrt{L}}\cos\big(\frac{\pi k}{L}\big)\sum_{j=1}^{L/4}\cos\big(\frac{\pi k}{(L/4)}j\big) -\frac{1}{\sqrt{L}}\sin\big(\frac{\pi k}{L}\big)\sum_{j=1}^{L/4}\sin\big(\frac{\pi k}{(L/4)}j\big) & \text{for } k \neq 0
    \end{cases}
\end{split}
\end{equation}
For $k \neq 0$, we evaluate the summations over $\cos$ and $\sin$ exactly, to get
\begin{equation}
\begin{split}    
    \sum_{j=1}^{L/4}\cos\bigg(\frac{\pi k}{(L/4)}j\bigg) &= \begin{cases}
        0 & \text{if  } k \in \mathrm{even} \\
        -1 & \text{if  } k \in \mathrm{odd}
    \end{cases}\\
    \sum_{j=1}^{L/4}\sin\bigg(\frac{\pi k}{(L/4)}j\bigg) &= \begin{cases}
        0 & \text{if  } k \in \mathrm{even} \\
        \cot(\frac{2\pi k}{L}) & \text{if  } k \in \mathrm{odd}.
    \end{cases}
\end{split}
\end{equation}
Therefore at $t=0$, for $k\in \mathrm{odd}$, the mode occupancy is given by 
\begin{equation}
\begin{split}
    \Tilde{G}_k(\tau=0) &= \frac{1}{\sqrt{L}}\cos\bigg(\frac{\pi k}{L}\bigg) - \frac{1}{\sqrt{L}}\sin\bigg(\frac{\pi k}{L}\bigg)\frac{\cos(2\pi k/L)}{\sin(2\pi k/L)} \\
    &= \frac{1}{\sqrt{L}}\cos\bigg(\frac{\pi k}{L}\bigg) - \frac{1}{\sqrt{L}}\sin\bigg(\frac{\pi k}{L}\bigg)\frac{\cos(2\pi k/L)}{2\sin(\pi k/L)\cos(\pi k/L)} \\
    &= \frac{1}{\sqrt{L}}\left(\frac{2\cos^2(\pi k/L) - \cos(2\pi k/L)}{2\cos(\pi k/L)} \right) \\
    &= \frac{1}{\sqrt{L}}\left(\frac{2\cos^2(\pi k/L) - 2\cos^2(\pi k/L) + 1}{2\cos(\pi k/L)} \right) = \frac{1}{2\sqrt{L}}\frac{1}{\cos(\pi k/L)}.
\end{split}
\end{equation}
To summarize, the alternate initial condition of the bond length variables in the eigenbasis of $\tilde{\mathbf{U}}_{\mu\to\infty}$ can be expressed as 
\begin{equation}
    \Tilde{G}_k (\tau=0) = \begin{cases}
        -\sqrt{2L}/8 & \text{for  } k=0 \\
        \begin{cases}
            0 & \text{if  } k\in \mathrm{even} \\
            \frac{\sec(\pi k/L)}{2\sqrt{L}} & \text{if  } k\in \mathrm{odd}
        \end{cases} & \text{for  } k>0.
    \end{cases}
\end{equation}
Using this result, the bond length variables at time $\tau=2\Lambda t$ can be obtained as 
\begin{equation}
\begin{split}
    G_p(\tau) &= \sum_{k=0}^{L/2-1} \mathcal{F}^{-1}_{pk} \,e^{\lambda^{(k)}\tau}\,\Tilde{G}_k(\tau=0) \\
    &= -\frac{1}{4} + \frac{1}{\sqrt{L/4}}\sum_{k=1}^{L/2-1}\cos\bigg(\frac{\pi k(2p-1)}{L}\bigg) \, e^{\lambda_k\tau}\, \Tilde{G}_k(\tau=0) \\
    &= -\frac{1}{4} + \frac{1}{L}\sum_{k\in \mathrm{odd}}\cos \bigg(\frac{\pi k(2p-1)}{L}\bigg)\,e^{\lambda_k\tau}\,\sec\bigg(\frac{\pi k}{L}\bigg)\\
    &= -\frac{1}{4} + \frac{1}{L}\sum_{k\in \mathrm{odd}}\left(\cos\bigg(\frac{2\pi p k}{L}\bigg) + \tan\bigg(\frac{\pi k}{L}\bigg)\sin\bigg(\frac{2\pi p k}{L}\bigg)\right)\,e^{\lambda_k\tau}.
\end{split}
\end{equation}
\textit{Late-time approximation:} At sufficiently late times, low-lying modes ($k\ll L$) dominate the dynamics (see Eq.~\eqref{eq:eigenpairs}) and higher modes $k\sim \mathcal{O}(L)$ decay to zero. Hence, we take (i) $\lambda^{(k)}\approx -4\pi^2k^2/L^2$, (ii) $\tan(\pi k/L)\approx \pi k/L$ and obtain 
\begin{equation}
\begin{split}
    G_p(t) &\approx -\frac{1}{4} + \frac{1}{L}\sum_{k\in\mathrm{odd}}\left(\cos\bigg(\frac{2\pi p k}{L}\bigg) + \frac{\pi k}{L}\sin\bigg(\frac{2\pi p k}{L}\bigg)\right)e^{-\frac{4\pi^2k^2}{L^2}\tau} \\
    &\approx -\frac{1}{4} + \frac{1}{L}\sum_{k\in\mathrm{odd}}\cos\bigg(\frac{2\pi p k}{L}\bigg)e^{-\frac{4\pi^2k^2}{L^2}\tau} =  -\frac{1}{4} + H_p(\tau)
\end{split}.
\end{equation}
where the second term we introduce as $H_p(t)$ and write it as 
\begin{equation}
\begin{split}
    H_p(\tau) &= \frac{1}{L}\sum_{k\in \mathrm{odd}}\cos\bigg(\frac{2\pi p k}{L}\bigg)e^{-\frac{4\pi^2k^2}{L^2}\tau} \\
    &= \frac{1}{L}\sum_{k=0}^{L/2-1}\cos\bigg(\frac{2\pi p k}{L}\bigg)e^{-\frac{4\pi^2k^2}{L^2}\tau} - \frac{1}{L}\sum_{k=0}^{L/2-1}\cos\bigg(\frac{2\pi p (2k)}{L}\bigg)e^{-\frac{4\pi^2(2k)^2}{L^2}\tau}.
\end{split}
\end{equation}
In the limit $L\to\infty$ and $p\ll L$, the summations above can be approximated as integrals. We then obtain 
\begin{equation}
\begin{split}
   \frac{1}{L}\sum_{k=0}^{L/2-1}\cos\bigg(\frac{2\pi p k}{L}\bigg)e^{-\frac{4\pi^2k^2}{L^2}\tau} &\approx \frac{1}{2\pi}\int_0^{\infty}dx \cos(px)e^{-x^2\tau} = \frac{1}{4\sqrt{\pi \tau}}e^{-p^2/4\tau} \\
   \frac{1}{L}\sum_{k=0}^{L/2-1}\cos\bigg(\frac{2\pi p (2k)}{L}\bigg)e^{-\frac{4\pi^2(2k)^2}{L^2}\tau} &\approx \frac{1}{4\pi}\int_0^{\infty}dx \cos(px)e^{-x^2\tau} = \frac{1}{8\sqrt{\pi \tau}}e^{-p^2/4\tau},
\end{split}
\end{equation}
which gives us,
\begin{equation}
\begin{split}
    G_p(\tau)
        &\approx -\frac{1}{4} + \frac{1}{8\sqrt{\pi \tau}}e^{-p^2/4\tau}, \qquad \text{ for } p\ll L.
\end{split}
\end{equation}
By definition of the bond length $p=d(m,n)$, as discussed in Sec.~(\ref{sec:supp_bond_length_derivation}), we finally obtain the expression for the diagonals of the four-point correlators in the thermodynamic limit $L\to\infty$ as 
\begin{equation}
\label{eq:supp_4pt_profile}
    F_{mnmn}(t) = {\rm Tr}\big[\hat{c}^{\dagger}_m \hat{c}^{\dagger}_n \hat{c}_m\hat{c}_n\hat{\rho}(t)\big] \approx -\frac{1}{4} + \frac{1}{8\sqrt{4\pi\Lambda t}}e^{-|m-n|^2/(16\Lambda t)}.
\end{equation}
Where $|m-n|$ denotes the distance between the lattice sites $m$ and $n$. Recall that $\Lambda= 2 J^2/\gamma$ is the diffusion constant. $F_{mnmn}(t)$ has a Gaussian form and features translational symmetry, as discussed in the main text. The above expression matches with Eq.~\eqref{eq:two_particle_profile} in the main text.

\subsection{Diffusive growth of particle number fluctuation for tight-binding ($\mu\to\infty$) lattice setup}
We now use the expression for $F_{mnmn}(t)$ in  Eq.~\eqref{eq:supp_4pt_profile} to analytically establish the diffusive growth of particle number fluctuations $\sigma_{M}^2(t)=\langle(\sum_{i=0}^{M-1}\hat{n}_i)^2\rangle-\langle\sum_{i=0}^{M-1}\hat{n}_i\rangle^2$ in a domain of size $M \ll L$.  $\sigma_M^2(t)$ can be expressed in terms of the bond length variables as,  
\begin{equation}\label{eq:SR_in_bond_len}
\begin{split}
    \sigma_M^2(t) &= -2\sum_{p=1}^{M}\left(M-p\right)G_p(t) + M/2 - (M/2)^2 \\
        &= -2\sum_{p=1}^{M}\left(M-p\right)\left(-\frac{1}{4}+H_p(t)\right) + M/2 - (M/2)^2
\end{split}
\end{equation}
The summation over $H_p$ can be carried out as follows:
\begin{equation}
\begin{split}
    &\sum_{p=1}^{M}\left(M-p\right)H_p(t) \approx M\sum_{p=0}^{M}\frac{e^{-p^2/4\Lambda t}}{8\sqrt{\pi \Lambda t}} - \sum_{p=0}^{M}p\frac{e^{-p^2/4\Lambda t}}{8\sqrt{\pi \Lambda t}} - \frac{M}{8\sqrt{\pi t}}
\end{split}
\end{equation}
Considering $\Lambda t\gg 1$, the above summation can be approximated as the following integrals,
\begin{equation}
\begin{split}
    \sum_{p=1}^{M}\left(M-p\right)H_p(t) &\approx\frac{M}{8\sqrt{\pi}}\int_0^{M/\sqrt{\Lambda t}}dx e^{-x^2/4} - \frac{\sqrt{\Lambda t}}{8\sqrt{\pi}}\int_0^{M/\sqrt{\Lambda t}}dx\, x\cdot e^{-x^2/4}  - \frac{M}{8\sqrt{\pi \Lambda t}}
\end{split}
\end{equation}
which upon taking the limit $M/\sqrt{\Lambda t}\gg 1$, i.e., $t \ll M^2/\Lambda$ gives,
\begin{equation}
    \sum_{p=1}^{M}\left(M-p\right)H_p(t) \approx \frac{M}{8} - \sqrt{\frac{\Lambda t}{16 \pi}}
\end{equation}
Substituting the above in the expression for domain number fluctuation in Eq.~\eqref{eq:SR_in_bond_len} gives,
\begin{equation}
\begin{split}
    \sigma^2_M(t) &= -2\sum_{p=1}^{M}\left(M-p\right)G_p(t) +M/2 - (M/2)^2\\
    &= \sum_{p=1}^{M}\frac{1}{2} -2\sum_{p=1}^{M}\left(M-p\right)H_p(t) + M/2 - (M/2)^2 \\
    &\approx M^2/4 - M/4 - M/4 + \sqrt{\Lambda t/(4\pi)} + M/2 - (M/2)^2\\
    &\approx \sqrt{\frac{\Lambda t}{4\pi}}
\end{split}
\end{equation}
Hence using the bond length representation we analytically show that for $ \frac{1}{\Lambda} \ll t \ll \frac{L^2}{\Lambda}$, the bipartite particle number fluctuations in a domain of size $M\ll L$ grow as $\sigma_{M}^2(t)\sim t^{\beta}$, with diffusive exponent $\beta=\frac{1}{2}$.
  
\section{Results for domain wall initial condition for long-range fermionic lattice}
In this section, we show the robustness of Family-Vicsek (FV) scaling of number fluctuation growth against different initial conditions. In the main text, we have chosen the alternating state $|\psi(0)\rangle = \prod_{m=0}^{L/2-1}c^{\dagger}_{2m} |0 \rangle$ does not have density variations at large scales. Here, we show the FV scaling of number fluctuation growth for an initial state with large scale density variations, which is the domain wall state, defined as $|\psi(0)\rangle = \prod_{m=0}^{L/2-1}c^{\dagger}_{m} |0 \rangle$.

\begin{figure}[!h]
    \centering
    \includegraphics[width=\linewidth]{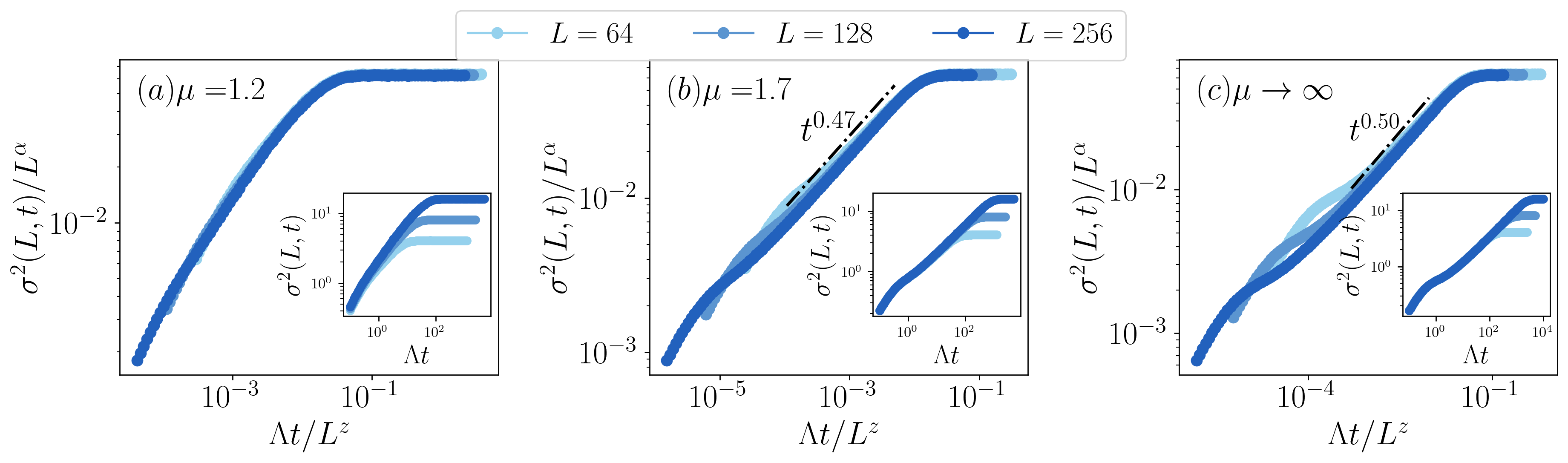}
    \caption{FV scaling of dynamics of bipartite particle number fluctuation $\sigma^2(L,t)$ for the long-range hopping lattice model, starting from the domain wall initial condition. The emergence of FV scaling behaviour is evident from the collapse in data upon scaling the ordinate and abscissa by $1/L^{\alpha}$ and $1/L^z$, respectively. We find super-diffusive exponents $(\alpha,\beta,z)= (1.0,1/(2\mu-1),2\mu-1$ for $\mu<1.5$ and diffusive scaling exponents  $(\alpha,\beta,z)= (1.0,0.5,2.0) $ for $\mu \geq 1.5$. The inset of the figures show unscaled dynamics for number fluctuation. Note that for the super-diffusive regime, (a) $\mu=1.2$, we fail to observe the $\sigma^2(L,t)\sim t^{\beta}$ scaling due to the finite-size effects in small lattice systems ($L\sim 10^2$).}
    \label{fig:FV_domain_wall}
\end{figure}

For such an initial state, we define the bipartite particle number fluctuations $\sigma^2(L,t)$ in the right half of the lattice, i.e., $\sigma^2(L,t)=\langle\hat{h}^2\rangle-\langle\hat{h}\rangle^2$ with $\hat{h}=\sum_{i=L/2}^{L-1}\hat{n}_i$. In Fig.~\ref{fig:FV_domain_wall} we show the FV scaling starting from the domain wall initial state, with scaling exponents $(\alpha,\beta,z)$ identical to the case of alternating initial state. That is, $(\alpha,\beta,z)=(1.0,1/(2\mu-1),2\mu-1)$ for $\mu<1.5$ and $(\alpha,\beta,z)=(1.0,0.5,2.0)$ for $\mu\geq 1.5$.

\section{Results for particle number fluctuation for bosons for dephased long-range lattice systems}

In this section, we provide results for the growth of number fluctuations when the lattice system consists of bosons. We discuss the case when the initial state is an alternating one, $|\psi(0)\rangle = \prod_{m=0}^{L/2-1}\hat{b}^{\dagger}_{2m} |0 \rangle$, and numerically show the Family-Vicsek (FV) scaling of number fluctuation growth for bosons in Fig.~\ref{fig:FV_bosons}. We obtain the same crossover from superdiffusive to diffusive regime with exactly same scaling exponents as was the case for fermions. 

The bosonic lattice system dynamics can be modelled by LQME:
\begin{equation}
\label{eq:master_eq_bosons}
    \frac{d}{dt} \hat{\rho} = -i\Big[\hat{H},{\hat{\rho}(t)}\Big] + \gamma\,\sum_{j=0}^{L-1} \Big(\hat{n}_j\,\hat{\rho}\,\hat{n}_j -\frac{1}{2}\{\hat{n}^2_j,\hat{\rho}\} \Big),
\end{equation}
where, the Hamiltonian is given as,
\begin{equation}
\label{eq:hamiltonian_bosons}
\hat{H} =  J\sum_{i=0}^{L-1} \sum_{j>i}^{L-1} \frac{1}{d(i,j)^\mu}\Big[\hat{b}_i\hat{b}_{i+r}^{\dagger} + \hat{b}_i^{\dagger}\hat{b}_{i+r}\Big],
\end{equation}
with $\hat{b}_i(\hat{b}^{\dagger}_i)$ being the bosonic annihilation (creation) operator at site $i$, and $\hat{n}_i=\hat{b}^{\dagger}_i\hat{b}_i$ is the bosonic number operator. $d(i,j)$ denotes the distance between lattice sites $i$ and $j$, which under periodic boundary conditions is given as $d(i,j)=\min(|i-j|,L-|i-j|)$.

Bipartite particle number fluctuation $\sigma^2(L,t)$ may be expressed as summations of diagonals of two-point and four-point correlators in the site basis, $\sigma^2(L,t)=\sum_{m,n=0}^{L/2-1}F_{mnmn} + \sum_{m=0}^{L/2-1}D_{mm}(1-\sum_{n=0}^{L/2-1}D_{nn})$. We first give the microscopic equations governing the dynamics of the two-point and four-point correlators, following which we obtain the effective equations for the same, under the strong dephasing limit ($\gamma \gg J$). We then develop the bond-length representation for the effective equations of four-point correlator.

\subsection{Microscopic equations for two-point and four-point correlation functions}
The microscopic equations of motion of $D_{mn}={\rm Tr}\big[ \hat{b}^{\dagger}_m\hat{b}_n\,\hat{\rho}\big]$ and $F_{mnpq}={\rm Tr}\big[\hat{b}^{\dagger}_m \hat{b}^{\dagger}_n \hat{b}_p\hat{b}_q\,\hat{\rho}\big]$, as obtained from the LQME in Eq.~\eqref{eq:master_eq_bosons} is given as,
\begin{equation}
\label{eq:2pt_micro_bosons}
\partial_t D_{mn} = iJ\sum_{j=\pm 1}^{\pm(L/2-1),L/2}\frac{D_{m(n+j)}-D_{(m+j)n}}{|j|^{\mu}}+ \gamma \, (\delta^m_n -1)D_{mn},
\end{equation}
and
\begin{equation}
\label{eq:4pt_micro_bosons}
\begin{split}
\partial_t F_{mnpq} =& +iJ\sum_{j=\pm 1}^{\pm(L/2-1),L/2} \frac{+F_{mn(p+j)q}+ F_{mnp(q+j)} - F_{(m+j)npq} - F_{m(n+j)pq}}{|j|^{\mu}} \\
                    & +\gamma(\delta^m_p + \delta^n_q + \delta^m_q + \delta^n_p - \delta^m_n - \delta^p_q - 2)F_{mnpq},
\end{split}
\end{equation}
respectively. Note that $(m+j)\equiv (m+j)\mod L$, and similarly for $(n+j),(p+j)$ and $(q+j)$.
\subsection{Effective equations for two-point correlation function}
In the strong dephasing limit $\gamma \gg J$, we adiabatically eliminate the coherences of two-point correlator, $D_{mn}$, to obtain effective equations amongst the diagonals $D_{mm}$. As explained in Sec. \ref{sec:supp_2pt_effective}, we assume that in time scale $\tau=Jt$ (i) $\partial_{\tau} D_{mn} \ll \frac{\gamma}{J}D_{mn}$ and (ii) $D_{mn}\ll D_{mm}$, to get:
\begin{equation}
\label{eq:2pt_effective_bosons}
    \partial_t D_{mm} = \frac{2J^2}{\gamma} \sum_{j\neq 0}\frac{D_{(m+j)(m+j)} - D_{mm}}{|j|^{2\mu}}
\end{equation}
Here, $(m+j)\equiv (m+j)\mod L$. Following the solution given in Sec.~\ref{sec:supp_2pt_effective_soln}, for alternating initial state, the above equation solves to
\begin{equation}
\label{eq:2pt_effective_bosons_soln}
    D_{mm}(t) = \frac{1}{2}\left(1 + e^{i\pi m}\,e^{-E(k=L/2)t}\right).
\end{equation}
where, the decay rate is:
\begin{equation}
    E(k=L/2) = \frac{8J^2}{\gamma}\sum_{j=1}^{\frac{L}{2}-1}\frac{\sin^2(\pi j/2)}{j^{2\mu}}.
    \label{dispersion_bosons}
\end{equation}
\subsection{Effective equations for four-point correlation function}
Here, we adiabatically eliminate the coherences of four-point correlator $F_{mnpq}$ from its microscopic equation Eq.~\ref{eq:4pt_micro_bosons}, assuming strong dephasing $\gamma \gg J$. Note that there are $({L \choose 2}+L)^2$ independent four-point correlators. Of these, ${L \choose 2}+L$ are the diagonals, of the form $F_{mnmn}$ or $F_{mnnm}$. Applying the assumptions and derivation of Sec.~\ref{sec:supp_4pt_effective} for bosons, we get the effective equation for the bosonic four-point correlator as,
\begin{equation}
\label{eq:4pt_effective_bosons}
\begin{split}
    \partial_t F_{mnmn} = \frac{2J^2}{\gamma}\sum_{j=\pm 1}^{\pm(L/2-1),L/2}\Bigg[&\frac{(1+\delta^m_n)F_{(m+j)n(m+j)n} - (1+\delta^{m+j}_n)F_{mnmn} }{|j|^{2\mu}} \\
    & + \frac{(1+\delta^m_n)F_{m(n+j)m(n+j)} - (1+\delta^{m}_{n+j})F_{mnmn} }{|j|^{2\mu}}\Bigg].  
\end{split}
\end{equation}
Here, $(m+j)\equiv (m+j)\mod L$ and $(n+j)\equiv (n+j)\mod L$.
\subsection{Bond length representation for  four-point correlation function for Bosons}
\begin{figure*}
    \centering
    \includegraphics[width=\linewidth]{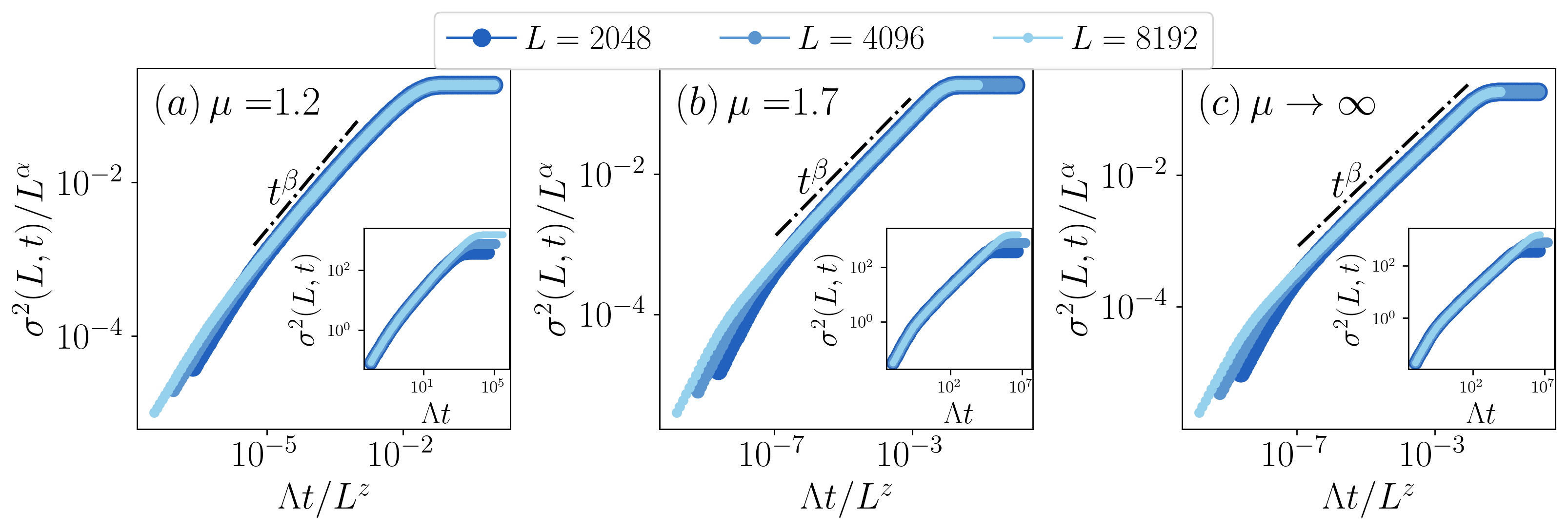}
    \caption{Plot for the growth and subsequent saturation of bipartite particle number fluctuation $\sigma^2(L,t)$ for the bosonic long-range hopping lattice model for different values of hopping exponent $\mu$. $\sigma^2(L,t)$ shows FV scaling with super-diffusive exponents $(\alpha,\beta,z)= (1.0,1/(2\mu-1),2\mu-1$ for $\mu<1.5$ and diffusive scaling exponents  $(\alpha,\beta,z)= (1.0,0.5,2.0) $ for $\mu \geq 1.5$, as evident from excellent the collapse in data for different lattice sizes $L$, upon scaling the ordinate and abscissa by $1/L^{\alpha}$ and $1/L^z$, respectively. The inset of the figures show unscaled dynamics for number fluctuation.}
    \label{fig:FV_bosons}
\end{figure*}

Bond length of a diagonal element of the four-point correlation matrix, $F_{mnmn}$ is defined as the distance between the sites $m$ and $n$. On a periodic lattice with $L$ sites, we have $p(m,n) = \min(|m-n|,L-|m-n|)$ and $p$ can takes values $0,1,\cdots,L/2$. Given the alternating initial condition, $|\psi(0)\rangle = \prod_{m=0}^{L/2-1}\hat{b}^{\dagger}_{2m} |0 \rangle$, we define the mapping from the diagonal elements $\{F_{mnmn} \}$ to the bond length variables $\{G^{(0),(1),(2)}_d \}$ as follows,
\begin{itemize}
    \item[--] Given a value $p\in\{1,3,..,L/2-1\}$, all diagonals $F_{mnmn}$ with bond length $d(m,n)=p$ are identical, and we label these as $G^{(0)}_p$. 
    That is, $F_{0101}=F_{1212} \cdots = F_{(L-1)0(L-1)0}=G_{1}^{(0)}$, $F_{0303}=F_{1414}= \cdots = F_{(L-1)2(L-1)2}=G_{3}^{(0)}$ and so on for $p=5,7,\cdots,L/2-1$
    \item[--] Given a value $p\in\{0,2,\cdots,L/2\}$, all diagonals $F_{mnmn}$ with (i) bond length $d(m,n)=p$ and (ii) both sites occupied in the initial state, are identical and we label these as $G^{(1)}_d$.
    That is, $F_{0000}=F_{2222}=\cdots=F_{(L-2)(L-2)(L-2)(L-2)}=G^{(1)}_0$, $F_{0202}=F_{2424}=\cdots=F_{(L-2)0(L-2)0}=G^{(1)}_2$ and so on for $p=4,6,\cdots,L/2$. 
    \item[--] Given a value $p\in\{0,2,\cdots,L/2\}$, all diagonals $F_{mnmn}$ with bond length $d(m,n)=p$ and both sites unoccupied in the initial state are identical, and we label these as $G^{(2)}_p$.
    That is, $F_{1111}=F_{3333}=\cdots=F_{(L-1)(L-1)(L-1)(L-1)}=G^{(2)}_0$, $F_{1313}=F_{3535}=\cdots=F_{(L-1)1(L-1)1}=G^{(2)}_2$ and so on for $p=4,6,\cdots,L/2$.
\end{itemize}
Applying this map to the effective equation of two-particle density matrix Eq.~\eqref{eq:4pt_effective_bosons} gives us the equation of motion of the bond length variables $\{G^{{(0)},{(1)},{(2)}}_p \}$,\\
For $p=0$:
\begin{equation}
\label{eq:bond_length_bosons1}
\begin{split}
    \partial_t G^{{(1)},{(2)}}_p = &2\Lambda\sum_{\substack{p' \in \mathrm{odd} \\ p' \neq p}}U_{pp'}\left(2G^{(0)}_{p'}-G^{{(1)},{(2)}}_p\right) + 2\Lambda\sum_{\substack{p' \in \mathrm{even} \\ p' \neq p}}U_{pp'}\left(2G^{{(1)},{(2)}}_{p'}-G^{{(1)},{(2)}}_p\right).
\end{split}
\end{equation}
For $p\neq 0$:
\begin{equation}
\label{eq:bond_lenght_bosons2}
\begin{split}
    \partial_t G^{(0)}_p = &2\Lambda\sum_{\substack{p' \in \mathrm{odd} \\ p' \neq p}}U_{pp'}\left(G^{(0)}_{p'}-(1+\delta^{p'}_0)G^{(0)}_p\right) + 2\Lambda\sum_{\substack{p' \in \mathrm{even} \\ p' \neq p}}U_{pp'}\left(\frac{G^{(1)}_{p'} + G^{(2)}_{p'}}{2}-(1+\delta^{p'}_0)G^{(0)}_p\right) \\
    \partial_t G^{{(1)},{(2)}}_p = &2\Lambda\sum_{\substack{p' \in \mathrm{odd} \\ p' \neq p}}U_{pp'}\left(G^{(0)}_{p'}-(1+\delta^{p'}_0)G^{{(1)},{(2)}}_p\right) + 2\Lambda\sum_{\substack{p' \in \mathrm{even} \\ p' \neq p}}U_{pp'}\left(G^{{(1)},{(2)}}_{p'}-(1+\delta^{p'}_0)G^{{(1)},{(2)}}_p\right).
\end{split}
\end{equation}
Here $\Lambda=2J^2/\gamma$ and the coupling matrix elements $U_{pp'}$ are given as:
\begin{equation}
    U_{pp'} = \begin{cases}
        \frac{1}{|{p'-p}|^{2\mu}} & \text{if  $p'=0,L/2$} \\
        \frac{1}{|p'-p|^{2\mu}} + 
        \begin{cases}
            1/(p+p')^{2\mu} & \text{if  $p+p' \leq L/2$} \\
            1/(L-(p+p'))^{2\mu} & \text{if  $p+p'>L/2$}
        \end{cases} & \text{if $0<p'<L/2$}
    \end{cases}
\end{equation}
For $p\in \mathrm{even}$, the occupied($G^{(1)}_p$) and unoccupied($G^{(2)}_p$) bonds contribute equally in the expression of surface roughness and the relevant quantity is $(G^{(1)}_p + G^{(2)}_p)/2$, for all $p\in \mathrm{even}$. Hence we use a representation where we average over the occupied and unoccupied bonds:
\begin{equation}
    G_p = \begin{cases}
        G^{(0)}_p & \text{for } p \in \mathrm{odd} \\
        (G^{(1)}_p + G^{(2)}_p)/2 & \text{for } p \in  \mathrm{even}
    \end{cases}
\end{equation}
Finally, we receive the bosonic bond length equations as
\begin{equation}
    \partial_t G_p = 2D\sum_{p'\neq p}U_{pp'}\left((1+\delta^p_0)G_{p'}-(1+\delta^{p'}_0)G_p\right).
\end{equation}
We use this bond length equations along with the solution for the diagonal elements for two-point correlator given in Eq.~\eqref{eq:2pt_effective_bosons} to compute particle number fluctuation for bosons in a domain. In Fig.~\ref{fig:FV_bosons} we show the emergence of FV scaling for bipartite number fluctuation growth for bosons. We find the same crossover from superdiffusive to diffusive regime with the same scaling exponents as was the case for fermions.
\end{document}